\documentclass[floatfix, twocolumn,amssymb,aps,prb,showpacs,superscriptaddress]{revtex4-2}
\usepackage{bm}
\usepackage{graphicx}
\usepackage{amsmath, amssymb}
\usepackage{braket}
\usepackage{color}
\usepackage{comment}
\begin{document}
\title{Photoinduced magnetic phase transitions in the cubic Kondo-lattice model}
\author{Ryo Hamano}
\author{Masahito Mochizuki}
\affiliation{Department of Applied Physics, Waseda University, Okubo, Shinjuku-ku, Tokyo 169-8555, Japan}
\date{Today}
\begin{abstract}
We theoretically study photoinduced magnetic phase transitions and their dynamical processes in the Kondo-lattice model on a cubic lattice. It is demonstrated that light irradiation gives rise to magnetic phase transitions from the ground-state ferromagnetic state to a three-dimensional antiferromagnetic state as a nonequilibrium steady state in the photodriven system. This phase transition occurs as a consequence of the formation of pseudo half-filling band occupation via the photoexcitation and relaxation of electrons, where all the electron states constituting the lower band separated from the upper band by an exchange gap are partially but nearly uniformly occupied. We also find that several types of antiferromagnetic correlations, e.g., A-type and C-type antiferromagnetic correlations, appear in a transient state of the dynamical phase transition. By calculating magnon spectra for the photodriven system, we argue that the instability to the A-type or C-type antiferromagnetic state occurs in the ferromagnetic ground state as a softening of the magnon band dispersion at corresponding momentum points depending on the light polarization. Our findings provide important insights into the understanding of photoinduced magnetic phase transitions in the three-dimensional Kondo-lattice magnets.
\end{abstract}
\maketitle

\section{Introduction}
The interaction of light and electrons in materials enables us to change and manipulate material states by light irradiation~\cite{Kirilyuk2010,Basov2011,Aoki2014,Mentink2017,Barman2020,CWang2020,dlTorre2021}. Many of the states of matters realized by light irradiation through the light-electron coupling are unique to nonequilibrium systems and often host various interesting physical phenomena and useful material functions. In addition, the control of states of matters with light is superior to other methods in terms of fast speed, energy saving, and non-abrasiveness. Therefore, with the development of laser technology, research on the control of material properties with light has become of increasing interest from the viewpoints of both fundamental science and technical applications.

In particular, the control of magnetism with light and photoinduced magnetic phase transitions have been studied intensively both experimentally~\cite{Beaurepaire1996,Koshihara1997,Fiebig1998,Averitt2001,Rini2007,Satoh2010,Mertelj2010,Radu2011,Ichikawa2011,Zhao2011,Yada2016,Razdolski2017,Lin2018,Mikhaylovskiy2020,Je2018,Ogawa2015,Gerlinger2021} and theoretically~\cite{Ono2017,Ono2018,Ono2019,Inoue2022,Hattori2024,Koshibae2009,Koshibae2011,Matsueda2007,Kanamori2009,Ohara2013,Mentink2015,Chovan2006,PWang2020,Stepanov2017,Fujita2017,Yang2018,Matsyshyn2023} as techniques and phenomena directly related to the spintronic applications. The Kondo-lattice models, in which localized spins and itinerant electrons are strongly coupled via exchange interactions~\cite{Zener1951,Anderson1955,Gennes1960}, are important targets of the research in this field because of their high potential for efficient magnetic manipulation via electronic excitations~\cite{Ono2017,Ono2018,Ono2019,Inoue2022,Hattori2024,Koshibae2009,Koshibae2011}. Usually, the energy scale of coupling between the light magnetic field and the magnetizations in magnetic materials through Zeeman interactions is very small. On the other hand, the energy scale of coupling between the light electric field and the electron charges is two to three orders of magnitude larger. As a result, in systems in which charge and magnetism are strongly coupled, e.g., Kondo-lattice magnets~\cite{Hayami2021R,Kawamura2025R}, Rashba electron systems~\cite{Tanaka2020a,Mochizuki2018,Shimizu2020}, and multiferroics~\cite{Katsura2005,Katsura2007,Mochizuki2010a,Mochizuki2010b,Mochizuki2011,Mochizuki2024}, it is expected that the charge excitations activated by the light electric field can efficiently manipulate, excite, and switch the magnetism via the light-induced charge excitations.

Therefore, various theoretical studies of the photoinduced magnetic phase transitions and the photoinduced magnetism have been performed for Kondo-lattice models. For example, in Ref.~\cite{Ono2017}, Ono and Ishihara theoretically demonstrated that a phase transition from ferromagnetism to antiferromagnetism occurs when the Kondo-lattice model on a square lattice is irradiated by laser light. They revealed a photoinduced destabilization of the ferromagnetic order by calculating magnon dispersion relations under the light irradiation using the nonequilibrium Green's function method~\cite{Ono2018}. The existence of transient processes depending on the light polarization was also discussed in this study.

Subsequently, Inoue and Mochizuki theoretically studied the photoinduced magnetic phase transitions in the Kondo-lattice model on a triangular lattice~\cite{Inoue2022}. They found that the ground-state ferromagnetic state changes to the 120-degree antiferromagnetic state when the system is irradiated by laser light. This 120-degree spin phase is known to be a ground state of this model at half filling in equilibrium~\cite{Akagi2010,Akagi2012,Azhar2017}. By investigating the dynamical process of this photoinduced phase transition in detail, they revealed that the pseudo half-filling band occupation due to the photoelectron excitations and relaxation as well as the photoinduced band modulation work as a universal mechanism that governs long-range magnetic orders under the light irradiation.

Theoretical studies on the photoinduced magnetic phase transitions in Kondo-lattice models have been limited to two-dimensional systems so far~\cite{Ono2017,Ono2018,Ono2019,Inoue2022,Hattori2024,Koshibae2009,Koshibae2011}. However, the real materials investigated in experiments are usually three dimensional. It is also known that the critical properties and behaviors of phase transitions at equilibrium strongly depend on the dimensions of the system. Therefore, for a comprehensive understanding of the photoinduced magnetic phase transitions in the spin-charge coupled systems such as Kondo-lattice models, it is of great importance to study the three-dimensional systems.

In this paper, we theoretically study nature and properties of the photoinduced magnetic phase transitions in a three-dimensional spin-charge coupled system by taking the Kondo-lattice model on a cubic lattice as the most fundamental example. Our numerical simulations show that laser irradiation on the model with a ferromagnetic ground state results in emergence of a three-dimensional antiferromagnetic order called G-type antiferromagnetism as a nonequilibrium steady state in the photodriven system. It is found that in the process of this dynamical phase transition, a certain transient state with a low-dimensional antiferromagnetic correlation called A-type or C-type antiferromagnetic correlation appears, type of which depends on the polarization of light. Furthermore, by calculating the magnon spectra using the nonequilibrium Green's function method, we reveal that the light irradiation first induces a softening of the magnon band dispersion at momentum points corresponding to the magnetic wavenumbers of the A-type or C-type antiferromagnetism, which destabilizes the ferromagnetic ground state. The findings of this work are expected to provide important insights into comprehensive understanding of the photoinduced magnetic phase transitions and establishment of techniques for optical manipulation of magnetism via charge excitations in spin-charge coupled systems.

The remaining part of this paper is structured as follows. In Sec.~II, we introduce a theoretical model and a method of the numerical simulations used in the present work. More specifically, we explain the Kondo-lattice model on a cubic lattice with localized spins coupled to electrons via the exchange coupling and the numerical method in which time-evolution equations for the localized spins and the electrons are simultaneously solved. In Sec.~III, we discuss photoinduced magnetic phase transitions from ferromagnetic to three-dimensional antiferromagnetic states based on the simulated results. We discuss dynamical processes of the phase transitions by focusing on calculated time profiles of spin structure factors, densities of states, band structures, and electron occupations. We reveal that several different types of antiferromagnetic short-range correlations, e.g., A-type and C-type antiferromagnetic correlations, appear in a transient state during the process of the dynamical phase transition, type of which depends sensitively on the light polarization. Also, we discuss the photoinduced instability of the ferromagnetic ground state towards a certain type of antiferromagnetic state caused by the softening of the magnon band dispersion by calculating the magnon spectra using the theory. Section~IV is devoted to the summary and conclusion.

\section{Formulation}
\subsection{Unit conversions}
In this paper, we use the transfer integral $t=$1 eV and the lattice constant $a$=5 \AA\; as the units of energy and length, respectively, and adopt natural units with $e=\hbar=1$. The unit conversions from dimensionless frequency $\omega$, electric field $E$, and time $\tau$ to their real values $\tilde{\omega}$, $\tilde{E}$ and $\tilde{\tau}$ in SI units are given as, 
\begin{itemize}
\item Frequency of light\\
\hspace{0.5cm}
$\displaystyle 
\omega =\frac{\hbar\tilde{\omega}}{t}=1
\quad \Leftrightarrow \quad
\frac{\tilde{\omega}}{2\pi} \approx 242$ [THz],
\item Amplitude of the light electric field\\
\hspace{0.5cm}
$\displaystyle 
E=\frac{e a \tilde{E}}{t}=1
\quad \Leftrightarrow \quad
\tilde{E} \approx 20$ [MV/cm],
\item Time\\
\hspace{0.5cm}
$\displaystyle 
\tau=\frac{\tilde{\tau}t}{\hbar}=1
\quad \Leftrightarrow \quad
\tilde{\tau}\approx 0.66$ [fs].
\end{itemize}

\subsection{Model}
We start with a time-dependent Hamiltonian for the photodriven Kondo-lattice model on a cubic lattice,
\begin{align}
\label{eq:tH0}
\mathcal{H}(\tau)=\mathcal{H}_{\rm kin}(\tau) + \mathcal{H}_{\rm exc}(\tau),
\end{align}
with
\begin{align}
\label{eq:tH1}
&\mathcal{H}_{\rm kin}(\tau)=\sum_{\langle i,j\rangle, s} 
t_{ij}(\tau)\; c_{is}^\dag c_{js}+\text{h.c.}
\\
\label{eq:tH2}
&\mathcal{H}_{\rm exc}(\tau)=-\frac{J}{2S}\sum_{i,s,s'}
\bm{S}_i(\tau) \cdot c_{is}^\dag \bm{\sigma}_{ss'} c_{is'}.
\end{align}
Here h.c. denotes the Hermitian conjugate, $c_{is}^\dag$ ($c_{is}$) is a creation (annihilation) operator of an electron with spin $s$ on the $i$th lattice site, and $t_{ij}(\tau)$ denotes the time-dependent transfer integrals between the nearest neighbor lattice sites at time $\tau$. The time-dependent localized spins $\bm S_i(\tau)$ are treated as classical vectors whose norm is $S$, and $\bm \sigma_{ss'}=(\sigma_{ss'}^{x},\sigma_{ss'}^{y},\sigma_{ss'}^{z})$ are the Pauli matrices. The first term describes the kinetic energies of the electrons, and the second term describes the exchange coupling between the localized spins and the electron spins, whose strength is given by the Kondo-coupling constant $J$. 

The Kondo-lattice model was originally proposed to describe heavy-fermion systems exhibiting the Kondo effect, where quantum localized spins are coupled antiferromagnetically to conduction electron spins. However, in recent years, there has been considerable research on the Kondo-lattice model with classical localized spins~\cite{Hayami2021R,Kawamura2025R}. In addition, real magnetic materials with large spins, which realize the Kondo-lattice model with classical spins, have been experimentally discovered and synthesized~\cite{Kawamura2025R}. It is important to note that the Kondo-lattice model with classical localized spins exhibits identical behaviors regardless of whether the exchange interaction is ferromagnetic or antiferromagnetic. In the present work, we consider a ferromagnetic coupling for the exchange interaction $J$, in which case the model is commonly referred to as the double-exchange model. This model has been extensively studied as a suitable framework for perovskite manganese oxides that exhibit colossal magnetoresistance effects. Given that we adopt the ferromagnetic exchange interaction, our model can also be described as a double-exchange model. However, we continue to refer to it as the Kondo-lattice model to explicitly indicate that the results and the proposed physics are independent of the sign of the exchange interaction. In this work, we explore the Kondo-lattice model on a cubic lattice. Therefore, from the perspective of crystal structure, transition-metal oxides with a three-dimensional perovskite structure, such as La$_{1-x}$Sr$_x$MnO$_3$ and SrFeO$_3$, are typical target materials. However, the phenomena we investigate in this study involve several universal aspects of physics that are independent of crystal structure. As a result, the target materials for this study are more broadly classified as Kondo-lattice magnets. For detailed introductions to various real Kondo-lattice magnets, readers are referred to Ref.~\cite{Kawamura2025R}.

Effects of the light irradiation are considered via the Peierls phases attached to the transfer integrals $t_{ij}$. In the present work, we consider only the transfer integrals between the nearest neighboring sites $i$ and $j$, which are given by,
\begin{align}
\label{eq:trf}
t_{ij}(\tau)= 
\begin{cases}
-t & (\tau < 0) \\
-t\exp[-i\bm{A}(\tau)\cdot (\bm{r}_i-\bm{r}_j)] & (\tau \ge 0).
\end{cases}
\end{align}
Here $\bm A(\tau)$ is a time-dependent vector potential of the light electromagnetic field. Note that the light electromagnetic field has, in principle, spatial dependence particularly in three-dimensional systems, where layers of atoms are stacked along the incident direction of light. We, however, assume that a uniform light electric field acts on the entire system and neglect the spatial dependence of the vector potential $\bm A(\tau)$. A typical frequency of the light considered in this work is $\Omega$=0.5, which corresponds to the wavelength of $\sim$2 $\mu$m. This wavelength is approximately 4000 times longer than the lattice constant of $a$=0.5 \AA. Therefore, as long as we focus on the area within this length scale, the above assumption is justified.

With the radiation gauge, the light electric field $\bm E(\tau)$ and the vector potential $\bm A(\tau)$ are related as,
\begin{align}
\label{eq:EEF}
\bm E(\tau)=-\frac{\partial \bm{A}(\tau)}{\partial \tau}.
\end{align}
In this study, we examine the following three cases with different light polarizations,\\
\begin{enumerate}
\item[1.] Linearly polarized light with $\bm E$$\parallel$$[100]$\\
\hspace{0.5cm}
$\displaystyle
\bm A(\tau)=A(\sin(\Omega\tau), 0, 0)$
\item[2.] Linearly polarized light with $\bm E$$\parallel$$[110]$\\
\hspace{0.5cm}
$\displaystyle 
\bm A(\tau)=A(\sin(\Omega\tau), \sin(\Omega\tau), 0)$
\item[3.] Circularly polarized light\\
\hspace{0.5cm}
$\displaystyle 
\bm A(\tau)=A(\cos(\Omega\tau), \sin(\Omega\tau), 0)$
\end{enumerate}

\begin{figure}[tb]
\includegraphics[scale=1.0]{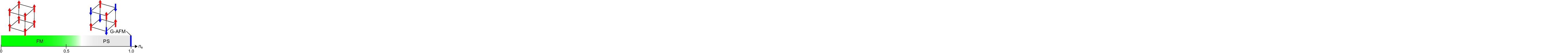}
\caption{Ground-state phase diagram of the cubic Kondo-lattice model for $J=14t$ as a function of the electron filling $n_{\rm e}$. Here FM, G-AFM, and PS denote the ferromagnetic phase, G-type antiferromagnetic phase, and phase separation, respectively. The FM phase appears when $0 < n_{\rm e} \le 0.3$, while the G-AFM state appears at $n_{\rm e}=0.5$. Spin configurations of the FM and G-AFM states are also shown.}
\label{Fig01}
\end{figure}
The ground-state phase diagrams of the cubic Kondo-lattice model have been studied in plane of the electron filling $n_{\rm e}$ and the Kondo coupling $J$~\cite{Henning2009,Hayami2014}. Here the electron filling $n_{\rm e}$ is defined by,
\begin{align}
\label{eq:ne}
n_{\rm e} \equiv \frac{N_{\rm e}}{2N}=\frac{1}{2N}\sum_{i,s}\braket{c^\dag_{is} c_{is}},
\end{align}
where $N_{\rm e}$ is the number of electrons in the system, and $N$ is the number of sites. In the following, we employ a system where $n_{\rm e}=0.25$ and $J=14t$, for which the ground-state spin configuration in the equilibrium system is ferromagnetic (FM), while the width of electron band $W$ is 12$t$. This FM state is stabilized by a physical mechanism called double-exchange mechanism associated with energy gains due to the electron transfers enhanced by the strong Kondo exchange coupling~\cite{Zener1951, Anderson1955,Gennes1960}. Starting from this initial FM state, we numerically simulate photoinduced dynamics of the localized spins and the electrons under the light irradiation. 

To discuss the ground-state band structure, we introduce the Fourier transformations of the electron operators and the spin operator, respectively, as, 
\begin{align}
\label{eq:ccdg}
c_{is}=\frac{1}{\sqrt{N}}\sum_{\bm k} c_{\bm ks} e^{i\bm k \cdot \bm r_i},
\quad
c_{is}^\dag=\frac{1}{\sqrt{N}}\sum_{\bm k} c_{\bm ks}^\dag e^{-i\bm k \cdot \bm r_i},
\end{align}
and
\begin{align}
\label{eq:SFr}
\bm S_i(\tau)=\frac{1}{\sqrt{N}}\sum_{\bm q} \bm S_{\bm q}(\tau) e^{i\bm q \cdot \bm r_i}.
\end{align}
Using these operators, we rewrite the Hamiltonian in the momentum representation as,
\begin{align}
\label{eq:HFr}
\mathcal{H}(\tau)=&\sum_{\bm k,s} \varepsilon_{\bm k+\bm A(\tau)} c_{\bm k s}^\dag c_{\bm k s}
\notag\\
&-\frac{J}{2S\sqrt{N}} \sum_{s,s'}\sum_{\bm k,\bm q}
\bm S_{\bm q}(\tau) \cdot c_{\bm k+\bm q s}^\dag \bm{\sigma}_{ss'} c_{\bm k s'}.
\end{align}
with
\begin{align}
\label{eq:ekA}
\varepsilon_{\bm k+\bm A(\tau)}=
-2t\sum_{\bm \mu=\hat{\bm x}, \hat{\bm y}, \hat{\bm z}} 
\cos \left\{(\bm k+\bm A(\tau)) \cdot \bm \mu \right\}.
\end{align}
In the FM state, all the localized spins are oriented along the $z$ direction for which $\bm S_{\bm q}$ takes a nonzero value only at $\Gamma$ point ($\bm q$=0) as $\bm S_{\bm q}=\sqrt{N}S\hat{\bm z}$. Hence, the Hamiltonian for the FM ground state is given by,
\begin{align}
\label{eq:HFM}
\mathcal{H}_{\rm FM}=\sum_{\bm k,s}\left\{
\varepsilon_{\bm k}-\frac{J}{2}{\rm sgn}(s) \right\}
c_{\bm k s}^\dag c_{\bm k s},
\end{align}
where the sgn function is defined as ${\rm sgn}(\uparrow)=+1$ and ${\rm sgn}(\downarrow)=-1$. This means that the initial state has two electron bands separated by the energy of $J$. For the present system with $n_{\rm e}$=0.25, the electron spins are perfectly polarized.

\subsection{Method}
We numerically simulate spatiotemporal dynamics of the electrons and the localized spins in the photodriven Kondo-lattice model described by the time-dependent Hamiltonian $\mathcal{H}(\tau)$ in Eqs.~\eqref{eq:tH0}-\eqref{eq:tH2}. The method used for numerical simulations is described below. The time-dependent Hamiltonian $\mathcal{H}(\tau)$ can be written in the matrix form using the orthonormal basis $\{\ket{\phi_{is}}\} \equiv \{c_{is}^\dag \ket{0}\}$ and satisfies the following eigenequation,
\begin{align}
\label{eq:egeq}
\mathcal{H}(\tau)\ket{\Psi_\nu(\tau)}=\varepsilon_\nu(\tau)\ket{\Psi_\nu(\tau)}.
\end{align}
Here $\nu$ ($=1,2, \cdots, 2N$) is an index of the eigenstates $\ket{\Psi_\nu(\tau)}$ in ascending order with respect to the corresponding eigenenergies $\varepsilon_\nu(\tau)$. The eigenstates $\{\ket{\Psi_\nu(\tau)}\}$ are orthonormal that satisfy the relation,
\begin{align}
\label{eq:orthn}
\sum_{\nu = 1}^{2N} \ket{\Psi_\nu(\tau)}\bra{\Psi_\nu(\tau)}=1.
\end{align}

We numerically simulate the one-particle wavefunctions $\ket{\tilde{\Psi}_\alpha(\tau)}$ starting with the ground-state wavefunctions, i.e., $N_{\rm e}$ eigenstates with lower eigenenergies of the static Hamiltonian before the light irradiation,
\begin{align}
\label{eq:iniwf}
\ket{\tilde{\Psi}_\alpha(0)}=\ket{\Psi_{\alpha}(0)} \hspace{1pc} (\alpha=1, 2, \cdots, N_{\rm e}).
\end{align}
The time evolutions of the one-particle wavefunctions $\ket{\tilde{\Psi}_\alpha(\tau)}$ for the light-irradiated system obey the following time-dependent Schr\"odinger equation,
\begin{align}
\label{eq:Scheq}
i\frac{\partial}{\partial \tau}\ket{\tilde{\Psi}_\alpha(\tau)}
=\mathcal{H}(\tau)\ket{\tilde{\Psi}_\alpha(\tau)}.
\end{align}
The time evolution is calculated by applying the time-evolution operator $\hat{U}(\tau,\tau+\Delta\tau)$ to the wavefunction as~\cite{Tanaka2022,Tanaka2020b,Tanaka2010},
\begin{align}
\label{eq:tev1}
\ket{\tilde{\Psi}_\alpha(\tau+\Delta\tau)}=
\hat{U}(\tau,\tau+\Delta\tau) \ket{\tilde{\Psi}_\alpha(\tau)},
\end{align}
where
\begin{widetext}
\begin{align}
\label{eq:tev2}
\hat{U}(\tau+\Delta\tau,\tau)
&=1+\sum_{n=1}^{\infty}(-i)^n \int_\tau^{\tau+\Delta\tau}d\tau_1
\int_\tau^{\tau_1}d\tau_2\;\cdots\; \int_\tau^{\tau_{n-1}}d\tau_n
\mathcal{H}(\tau_1) \;\cdots\; \mathcal{H}(\tau_n)
\\
&\approx \exp\left[-i\Delta\tau\mathcal{H}\left(\tau+\frac{\Delta\tau}{2}\right)\right]
\\
&\approx 1 -i\Delta\tau\;\mathcal{H}\left(\tau+\frac{\Delta\tau}{2}\right)
+\cdots +\frac{1}{8!}
\left[-i\Delta\tau\;
\mathcal{H}\left(\tau+\frac{\Delta\tau}{2}\right)
\right]^8.
\end{align}
The final form is used for simulations. We also adopt the following approximation,
\begin{align}
\label{eq:Happrx}
\mathcal{H}\left(\tau+\frac{\Delta\tau}{2}\right) \approx 
\frac{1}{2}\left\{
\mathcal{H}\left(\tau\right)+\mathcal{H}\left(\tau+\Delta\tau \right)
\right\}.
\end{align}
With this form of the approximated time-evolution operator, the computational errors are suppressed to be as small as the order of $\mathcal{O}(\Delta\tau^3)$.

Time evolutions of the localized spins $\bm S_i(\tau)$ are simulated using the Landau-Lifshitz-Gilbert (LLG) equation,
\begin{align}
\label{eq:LLGeq}
\frac{\partial \bm S_i(\tau)}{\partial \tau}
=\bm h_i^{\rm eff}(\tau) \times \bm S_i(\tau)
+\frac{\alpha_{\rm G}}{S} \bm S_i(\tau) \times \frac{\partial \bm S_i(\tau)}{\partial \tau},
\end{align}
where $\alpha_{\rm G}$ is the Gilbert-damping coefficient. This equation is numerically solved using the fourth-order Runge-Kutta method. The effective magnetic field $\bm h_i^{\rm eff}(\tau)$ is given by a derivative of the Hamiltonian $\mathcal{H}$ with respect to the localized spin $\bm S_i$ as,
\begin{align}
\label{eq:heff1}
\bm h_i^{\rm eff}(\tau)
&=-\left\langle \frac{\partial \mathcal{H}}{\partial \bm S_i} \right\rangle
=\frac{J}{2S}\sum_{s,s'} \left\langle c_{is}^\dag(\tau) \bm \sigma_{ss'} c_{is'}(\tau) \right\rangle.
\end{align}
A convenient expression of $\bm h_i^{\rm eff}(\tau)$ for the numerical simulations will be given later.

Starting with an initial set of the one-particle states $\ket{\tilde{\Psi}_\alpha(\tau=0)}=\ket{\Psi_\alpha(0)}$ ($\alpha=1, 2, \cdots, N_{\rm e}$) and an initial set of the localized spins $\{\bm S_i(\tau=0)\}$, we compute time evolutions of $\ket{\tilde{\Psi}_\alpha(\tau)}$ and $\{\bm S_i(\tau)\}$ by simultaneously solving Eq.~\eqref{eq:Scheq} and Eq.~\eqref{eq:LLGeq}. The wavefunction of the electron system at time $\tau$ is given by Cartesian products of the one-particle states as,
\begin{align}
\label{eq:wf1}
\ket{\tilde{\Phi}(\tau)} 
=\ket{\tilde{\Psi}_1(\tau)}\otimes\ket{\tilde{\Psi}_2(\tau)}\otimes \dots \otimes \ket{\tilde{\Psi}_{N_{\rm e}}(\tau)}.
\end{align}
In general, the one-particle states $\ket{\tilde{\Psi}_\alpha(\tau)}$, which evolve according to the time-dependent Schr\"odinger equation in Eq.~\eqref{eq:Scheq}, are different from the eigenstates $\ket{\Psi_\mu(\tau)}$ of the Hamiltonian at time $\tau$ obtained from the eigenequation in Eq.~\eqref{eq:egeq} except at $\tau=0$.

Using the eigenstates $\{\ket{\Psi_\nu(\tau)}\}$ as a set of the orthonormal basis, we introduce the following representation for the one-particle operator $\hat{O}(\tau)$,
\begin{align}
\label{eq:Oop1}
\hat{O}(\tau)=\sum_{\mu=1}^{2 N}\sum_{\nu=1}^{2 N}
\ket{\Psi_\mu(\tau)}\bra{\Psi_\mu(\tau)}{\hat{O}(\tau)}\ket{\Psi_\nu(\tau)}\bra{\Psi_\nu(\tau)}.
\end{align}
The expectation value of the operator $\hat{O}(\tau)$ is calculated by,
\begin{align}
\label{eq:Oop2}
\braket{\hat{O}(\tau)}
&=\braket{\tilde{\Phi}(\tau)|\hat{O}(\tau)|\tilde{\Phi}(\tau)}
\notag\\
&=\sum_{\alpha = 1}^{N_{\rm e}} \sum_{\mu,\nu=1}^{2N}
\braket{\tilde{\Psi}_\alpha(\tau)|\Psi_{\mu}(\tau)}
\braket{\Psi_{\mu}(\tau)|\hat{O}(\tau)|\Psi_{\nu}(\tau)}
\braket{\Psi_{\nu}(\tau)|\tilde{\Psi}_\alpha(\tau)}.
\end{align}
Using this equation, the electron occupation $n_\mu(\tau) $ of the $\mu$th eigenstate $\ket{\Psi_\mu(\tau)}$ and the total energy $E_{\rm tot}(\tau)$ per site are calculated by,
\begin{align}
\label{eq:nmu}
&n_\mu(\tau) \equiv 
\braket{\tilde{\Phi}(\tau)|\Psi_{\mu}(\tau)}
\braket{\Psi_{\mu}(\tau)|\tilde{\Phi}(\tau)}
=\sum_{\alpha=1}^{N_{\rm e}}\left|
\braket{\Psi_{\mu}(\tau)|\tilde{\Psi}_\alpha(\tau)}\right|^2,
\\
\label{eq:Etot}
&E_{\rm tot}(\tau) \equiv 
\frac{1}{N}\braket{\tilde{\Phi}(\tau)|\mathcal{H}(\tau)|\tilde{\Phi}(\tau)}
=\frac{1}{N} \sum_{\nu=1}^{2 N} \varepsilon_\nu 
\sum_{\alpha=1}^{N_{\rm e}}\left|
\braket{\Psi_{\nu}(\tau)|\tilde{\Psi}_\alpha(\tau)}\right|^2.
\end{align}
The expectation value of the one-particle operator $\hat{O}(\tau)$ is also given using another set of the orthonormal basis $\{\ket{\phi_{is}}\}$ as,
\begin{align}
\label{eq:Oop3}
\braket{\hat{O}(\tau)}
=\sum_{\alpha = 1}^{N_{\rm e}} \sum_{i,s}\sum_{j,s'}
\braket{\tilde{\Psi}_\alpha(\tau)|\phi_{is}}
\braket{\phi_{is}|\hat{O}(\tau)|\phi_{js'}}
\braket{\phi_{js'}|\tilde{\Psi}_\alpha(\tau)}.
\end{align}
Using this equation, the effective magnetic field $\bm h_i^{\rm eff}(\tau)$ given in Eq.~\eqref{eq:heff1} is rewritten in the form,
\begin{align}
\label{eq:heff2}
\bm h_i^{\rm eff}(\tau)
=\frac{J}{2S}\sum_{s,s'}\sum_{\alpha=1}^{N_{\rm e}} \bm \sigma_{ss'}
\braket{\phi_{is}|\tilde{\Psi}_\alpha(\tau)}^*
\braket{\phi_{is'}|\tilde{\Psi}_\alpha(\tau)}. 
\end{align}
This expression is used for numerical simulations. When we integrate the LLG equation using the fourth-order Runge-Kutta method to compute the time-evolved spins $\bm S_i(\tau+\Delta\tau)$, the effective magnetic fields at times $\tau+\Delta\tau/2$ and $\tau+\Delta\tau$ are required. They are calculated by,
\begin{align}
\label{eq:heff3}
\bm h_i^{\rm eff}(\tau+\Delta\tau)=\bm h_i^{\rm eff}(\tau)+\frac{d\bm h_i^{\rm eff}}{d\tau}\Delta\tau,
\quad
\bm h_i^{\rm eff}(\tau+\Delta\tau/2)=\bm h_i^{\rm eff}(\tau)+\frac{1}{2}\frac{d\bm h_i^{\rm eff}}{d\tau}\Delta\tau.
\end{align}
The time-derivative of $\bm h_i^{\rm eff}$ is given by the following formula, which is derived from Eq.~\eqref{eq:heff1} using Eq.~\eqref{eq:Scheq},
\begin{align}
\label{eq:dvheff}
\frac{d\bm h_i^{\rm eff}(\tau) }{d\tau}
&=\frac{J}{2S}\sum_{s,s'}\sum_{\alpha=1}^{N_{\rm e}}
\bm \sigma_{ss'}\left[ 
\braket{\phi_{is}|\frac{d}{d\tau}|\tilde{\Psi}_\alpha(\tau)}^*
\braket{\phi_{is'}|\tilde{\Psi}_\alpha(\tau)}
+
\braket{\phi_{is}|\tilde{\Psi}_\alpha(\tau)}^*
\braket{\phi_{is'}|\frac{d}{d\tau}|\tilde{\Psi}_\alpha(\tau)}\right]
\notag\\
&=\frac{iJ}{2S}\sum_{s,s'}\sum_{\alpha=1}^{N_{\rm e}}
\bm \sigma_{ss'}\left[ 
\braket{\phi_{is}|\mathcal{H}(\tau)|\tilde{\Psi}_\alpha(\tau)}^*
\braket{\phi_{is'}|\tilde{\Psi}_\alpha(\tau)}
-
\braket{\phi_{is}|\tilde{\Psi}_\alpha(\tau)}^*
\braket{\phi_{is'}|\mathcal{H}(\tau)|\tilde{\Psi}_\alpha(\tau)}\right]
\notag\\
&=-\frac{J}{S}\sum_{s,s'}\sum_{\alpha=1}^{N_{\rm e}}
\text{Im}\left[\bm \sigma_{ss'}
\braket{\phi_{is}|\mathcal{H}(\tau)|\tilde{\Psi}_{\alpha}(\tau)}^*
\braket{\phi_{is'}|\tilde{\Psi}_{\alpha}(\tau)} \right].
\end{align}
Here the relation $\bm \sigma_{s's}=\bm \sigma_{ss'}^*$ is used in the derivation.

To identify spatial structures of the localized spins, we calculate the spin structure factor $S(\bm q,\tau)$,
\begin{align}
\label{eq:Sq}
S(\bm q,\tau) \equiv \frac{1}{N^2S^2}\sum_{i, j=1}^N \bm S_i(\tau) \cdot \bm S_j(\tau)
e^{i\bm q \cdot (\bm r_i - \bm r_j)}.
\end{align}
\end{widetext}

\begin{figure}[tb]
\includegraphics[scale=1.0]{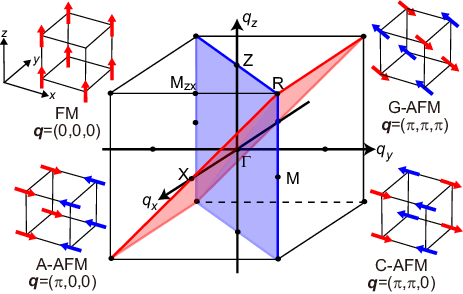}
\caption{Brillouin zone of the cubic lattice with high symmetric points. Spin configurations of the FM, A-AFM, C-AFM and G-AFM states are also shown, for which the spin structure factor $S(\bm q,\tau)$ has a peak in the momentum space at $\Gamma$ point [$\bm q=(0,0,0)$], X point [$\bm q=(\pi,0,0)$], M point [$\bm q=(\pi,\pi,0)$], and R point [$\bm q=(\pi,\pi,\pi)$], respectively.}
\label{Fig02}
\end{figure}
Note that $S(\bm q,\tau)$ has peaks at specific points in the momentum space depending on the magnetic state. Specifically, it has a peak at $\bm q=(0,0,0)$ for the FM state, at $\bm q=(\pi,0,0)$ for the A-AFM state with antiferromagnetic coupling along the $x$ axis, at $\bm q=(\pi,\pi,0)$ for the C-AFM state with antiferromagnetic $xy$ planes stacked ferromagnetically along the $z$ axis, and at $\bm q=(\pi,\pi,\pi)$ for the G-AFM state with three-dimensional antiferromagnetic coupling (see also Fig.~\ref{Fig02}).

\subsection{Simulation procedure}
The procedure of the numerical simulation is summarized in the following, which is equivalent to that used in the previous work~\cite{Inoue2022}. We first set an initial configuration of the localized spins $\{\bm S_i(\tau=0)\}$ to a nearly ferromagnetic state with small fluctuating components. Then we diagonalize the initial Hamiltonian $\mathcal{H}(\tau=0)$ constructed with $\{\bm S_i(\tau=0)\}$ to obtain the eigenvectors $\{\ket{\Psi_\mu(\tau=0)}\}$, which correspond to the one-particle wavefunctions $\{\ket{\tilde{\Psi}_\alpha(\tau=0)}\}$ at the ground state. Subsequently we repeat the following procedure in an iterative manner.
\begin{enumerate}
\item[1.] Calculate the effective magnetic fields $\bm h_i^{\rm eff}(\tau)$, $\bm h_i^{\rm eff}(\tau+\Delta\tau/2)$ and $\bm h_i^{\rm eff}(\tau+\Delta\tau)$ from $\{\ket{\tilde{\Psi}_\alpha(\tau)}\}$ using Eqs.~\eqref{eq:heff2}-\eqref{eq:dvheff}.\\

\item[2.] Calculate the localized spin vectors $\{\bm S_i(\tau+\Delta\tau)\}$ using the LLG equation in Eq.~\eqref{eq:LLGeq} with the above-calculated $\bm h_i^{\rm eff}(\tau)$, $\bm h_i^{\rm eff}(\tau+\Delta\tau/2)$ and $\bm h_i^{\rm eff}(\tau+\Delta\tau)$.\\

\item[3.] Calculate the time-dependent Hamiltonian $\mathcal{H}(\tau+\Delta\tau)$ in Eqs.~\eqref{eq:tH0}-\eqref{eq:tH2} with the above-calculated $\{\bm S_i(\tau+\Delta\tau)\}$.\\
 
\item[4.] Calculate the time-evolution operator $\hat{U}(\tau,\tau+\Delta\tau)$ in Eq.~\eqref{eq:tev2} by approximating $\mathcal{H}(\tau+\Delta\tau/2)$ with $\{\mathcal{H}(\tau)+\mathcal{H}(\tau+\Delta\tau)\}/2$.\\

\item[5.] Act the operator $\hat{U}(\tau,\tau+\Delta\tau)$ on the one-particle states $\{\ket{\tilde{\Psi}_\alpha(\tau)}\}$ to calculate $\{\ket{\tilde{\Psi}_\alpha(\tau+\Delta\tau)}\}$.
\end{enumerate}

The physical quantities are calculated at every time step using $\{\ket{\tilde{\Psi}_\alpha(\tau)}\}$ obtained by the time-evolution operation and $\{\ket{\Psi_\mu(\tau)}\}$ obtained by diagonalizing the Hamiltonian matrix $\mathcal{H}(\tau)$. The $2N \times 2N$ Hamiltonian matrix is a sparse matrix, which has only eight non-zero components in each row (column), i.e., two components associated with the on-site Kondo exchange coupling and six components associated with the transfer integrals on the nearest-neighbor bonds. We adopt the list mapping technique to reduce the computational cost in multiplying the matrix of the time-evolution operator to the one-particle wavefunctions. The computational cost for a simple multiplication of the Hamiltonian matrix is proportional to $8N^3$, while it is reduced to $16N(=8 \times 2N)$ when the list mapping method is used.

\section{Results}
\begin{table}[tbh]
\caption{Parameters used for the simulations}
\label{tab:params}
\begin{tabular}{l|l}
\hline
Physical quantity & Value \\
\hline \hline
Time step & $\Delta \tau$=0.005, 0.01\\
Transfer integral & $t=1$\\
Kondo coupling & $J=14$\\
Electron filling & $n_{\rm e}=0.25$ \\
Frequency of light & $\Omega=0.5$\\
Gilbert damping & $\alpha=0.5$\\
\hline
\end{tabular}
\end{table}
In this section, we will discuss the simulated results. We examine several cases of different light polarizations by taking the parameter values summarized in Table~\ref{tab:params} unless otherwise noted. 

Here, we discuss the choice of the parameter value $J=14t$. This value corresponds to perovskite manganese oxides, a typical class of materials described by our model. In the perovskite manganites, large localized spins with $S=3/2$, which behave as classical spins, are coupled to conduction-electron spins via the ferromagnetic exchange interaction originating from the Hund's-rule coupling, with a strength of $J_{\rm H} \approx 1$ eV. The transfer integral $t$ between neighboring Mn $e_g$-orbitals, mediated by the intervening oxygen $2p$-orbital, is approximately 0.20-0.25 eV. This implies that $J_{\rm H}$ is approximately $4t$-$5t$. On the other hand, by the definition of our Hamiltonian, the normalized coupling coefficient $J/2S$ corresponds to $J_{\rm H}$. Given that $S=3/2$, $J(=14t)$ corresponds to $J/2S=4.66t$, which reproduces the exchange interaction or Hund's-rule coupling $J_{\rm H}$ in the perovskite manganites.

We start the simulations with a FM spin configuration polarized in the $z$ direction as an initial state. However, the perfectly polarized FM state is not appropriate for numerical simulations because the light irradiation, in principle, cannot induce any spin-orientation changes without numerical rounding errors in this case. Hence, small random deviations from the $z$ axis are considered for the local spin orientations in the initial FM state, which are inevitably present at finite temperatures due to thermal fluctuations. Specifically, the initial spin configurations are given by,
\begin{align}
\label{eq:Svec}
\bm S_i(\tau=0)=(\sin\theta_i \cos\phi_i, \sin\theta_i \sin\phi_i, \cos\theta_i),
\end{align}
with
\begin{align}
\label{eq:Sfluc}
(\theta_i,\phi_i)=(m_i \delta\theta/N, 2\pi n_i/N),
\end{align}
where $m_i$ and $n_i$ are random integer numbers ranging from 1 to $N$. We set $\delta\theta$=0.012 throughout the following simulations, which corresponds to very small thermal fluctuations realized at an ultralow temperature of $k_{\rm B}T=0.0001t$  (see appendix~\ref{sec:Appendix3}). It is important to note that this temperature does not imply that the phenomena discussed in this work can only be observed at such ultralow temperatures. Rather, it indicates that only very small fluctuations are necessary to observe these phenomena.

\subsection{Linearly polarized light with $\bm E $$\parallel$[100]}
\begin{figure}[tb]
\includegraphics[scale=0.5]{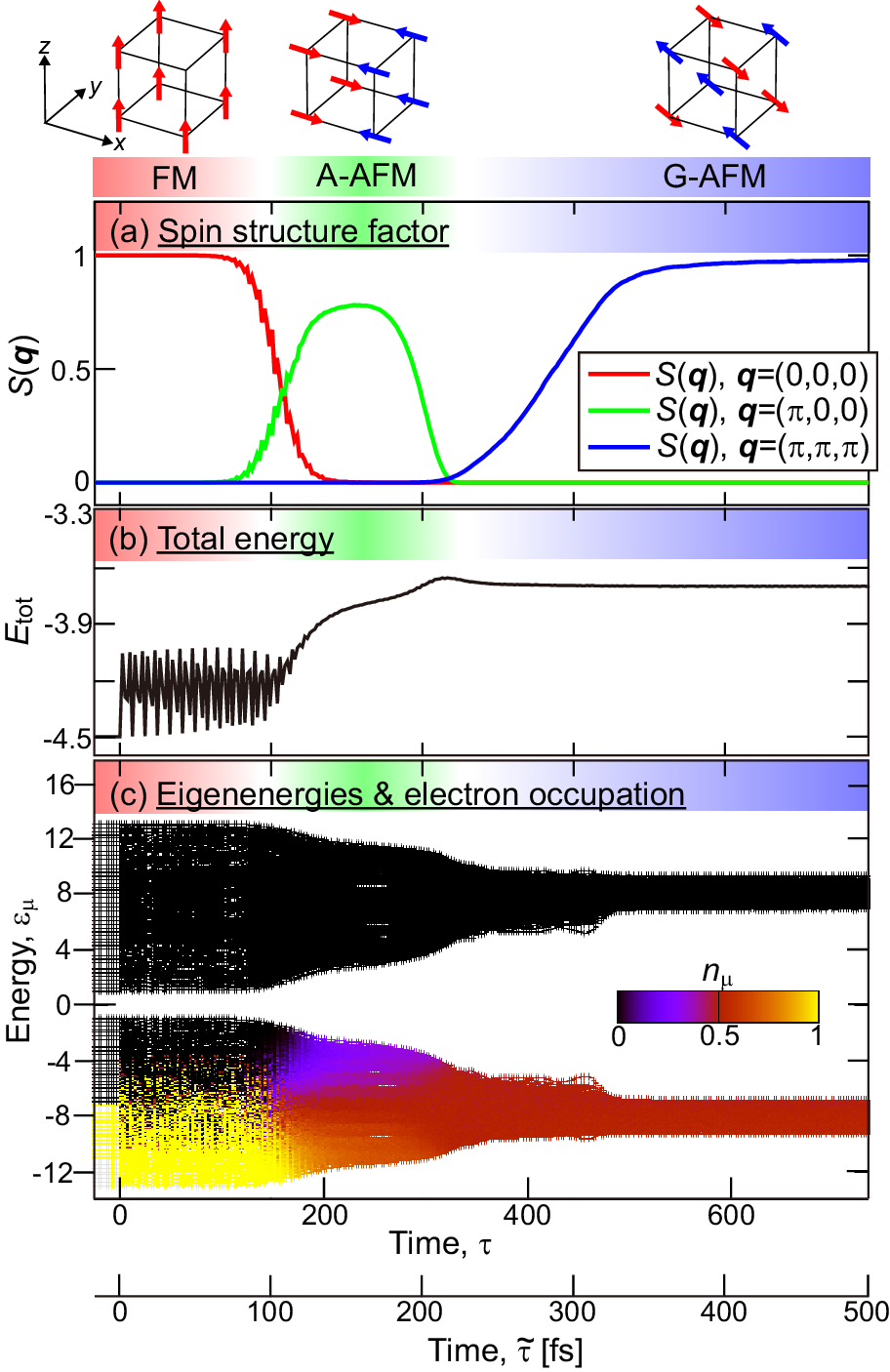}
\caption{Results for the linearly polarized light with $\bm E$$\parallel$[100], which indicate a photoinduced magnetic phase transition from FM to G-AFM phase via a transient phase with A-AFM spin correlations. (a) Time profiles of the spin structure factor $S(\bm q)$ at $\bm q=(0,0,0)$, $\bm q=(\pi,0,0)$ and $\bm q=(\pi,\pi,\pi)$, which quantify the FM, A-AFM and G-AFM spin correlations, respectively. (b) Time profile of the total energy. (c) Time profiles of the eigenenergies $\varepsilon_\mu$ of the time-dependent Hamiltonian $\mathcal{H}(\tau)$ and the electron occupation $n_\mu$. The simulations are performed for a system size of $N=12^3$ and the light field $\bm A(\tau)=A(\sin(\Omega\tau), 0, 0)$ with $A$=2.0 and $\Omega$=0.5. Here $\tilde{\tau}$ is the real time when $t$=1 eV.}
\label{Fig03}
\end{figure}
We first discuss results for the linearly polarized light with $\bm E$$\parallel$[100]. The simulations are performed for a system size of $N=12^3$ and the light field $\bm A(\tau)=A(\sin(\Omega\tau), 0, 0)$ with $A$=2.0 and $\Omega$=0.5. Figure~\ref{Fig03}(a) shows calculated time profiles of the spin structure factors $S(\bm q)$ at specific momentum points, $\bm q=(0,0,0)$, $\bm q=(\pi,0,0)$ and $\bm q=(\pi,\pi,\pi)$, which quantify the FM, A-AFM and G-AFM spin correlations, respectively. Here the A-AFM is an antiferromagnetic state with spins aligned antiferromagnetically along one axis ($x$ axis in the present case) and ferromagnetically along other two axes ($y$ and $z$ axes), while the G-AFM is an antiferromagnetic state with spins aligned antiferromagnetically along all the $x$, $y$ and $z$ directions. This figure indicates that a photoinduced magnetic phase transition from FM to G-AFM phases occurs via a transient phase with strong A-AFM spin correlations. 

Figure~\ref{Fig03}(b) shows the calculated time profile of the total energy $E_{\rm tot}$ of the electron system, which gradually increases when the A-AFM correlation quantified by $S(\bm q)$ at $\bm q=(\pi,0,0)$ begins to grow. This increase of $E_{\rm tot}$ is caused by the growth of antiferromagnetic correlations, which suppress the gain of kinetic energies of electrons due to the double exchange mechanism. The spin structure factor $S(\bm q)$ at $\bm q=(\pi,0,0)$ begins to decrease after taking a maximum value and is suppressed to zero. Subsequently, the G-AFM correlations, i.e., $S(\bm q)$ at $\bm q=(\pi,\pi,\pi)$, start to increase. The total energy $E_{\rm tot}$ continues to increase until $S(\bm q)$ at $\bm q=(\pi,0,0)$ is suppressed to zero and becomes almost constant after $S(\bm q)$ at $\bm q=(\pi,\pi,\pi)$ begins to rise and the system reaches the G-AFM phase.

Figure~\ref{Fig03}(c) shows calculated time evolutions of the eigenenergies $\varepsilon_\mu$ (dots) obtained by diagonalizing the time-dependent Hamiltonian $\mathcal{H}(\tau)$ in Eq.~\eqref{eq:HFr} and the electron occupation of the corresponding eigenstates $n_\mu$ (colors). The energy band is constituted with two portions, i.e., upper and lower bands, whose band centers are separated by the exchange energy of $J(=14t)$. Before the photoirradiation, both bands has an energy width of $W(=12t)$, and thus there is a gap of $\Delta(=2t)$ between the upper and lower bands, which render the system insulating. The electrons occupy a quarter of the eigenstates from the lowest energy. 

Right after the photoirradiation starts ($0< \tau \lesssim 180$), $S(\bm q)$ at $\bm q=(0,0,0)$ remains to be nearly unity, which indicates that the FM state remains stable. During this time, there is no noticeable change in the bandwidth for both the upper and lower bands. The electron occupation is also limited to the relatively low-energy states of the lower band. However, after $\tau \approx 200$, the bandwidth begins to decrease, and the excited electrons also begin to occupy the higher-energy states of the lower band. Around this change in the band structure and the electron occupation, the FM correlation, i.e., $S(\bm q)$ at $\bm q=(0,0,0)$,  begins to be suppressed and the A-AFM correlation, $S(\bm q)$ at $\bm q=(\pi,0,0)$, starts to grow alternatively. This bandwidth reduction is caused by the so-called dynamical localization effect or dynamical band contraction in photoexcitation~\cite{Dunlap1986,Grossmann1991,Ishikawa2014,Holthaus1992,Kayanuma2008,Eckardt2005,Lignier2007,Ohmura2021}. 

In the photodriven tight-binding models, the Peierls phases associated with the time-dependent electromagnetic vector potential $\bm A(\tau)$ are attached the transfer integrals $t_{ij}$. According to the Floquet theory, the transfer integrals in such systems are effectively renormalized by a factor of the Bessel function $J_0(\mathcal{A}_{ij})$ where $\mathcal{A}_{ij}=e{\bm E}^\omega \cdot (\bm r_i-\bm r_j)/\hbar\omega$~\cite{Yonemitsu2017,Kitayama2020,Kitayama2021a,Kitayama2021b,Kitayama2022}. Consequently, the bandwidth should be suppressed by the photoirradiation. Because the band gap between the upper and lower bands is determined by the relative magnitudes of the bandwidth and  the exchange-splitting due to the Kondo coupling, the gap becomes larger in the photoirradiated systems. Specifically, the linearly polarized light electric field in the $x$ direction effectively suppresses the transfer integrals on bonds along the $x$ direction~\cite{Kitayama2020,Kitayama2021a,Kitayama2021b,Kitayama2022}. This band contraction increases the Fermi-surface nesting in the $k_x$ direction and eventually enhances the antiferromagnetic correlation in the $x$ direction, resulting in the emergence of A-AFM correlation.

\begin{figure*}[tb]
\includegraphics[scale=1.0]{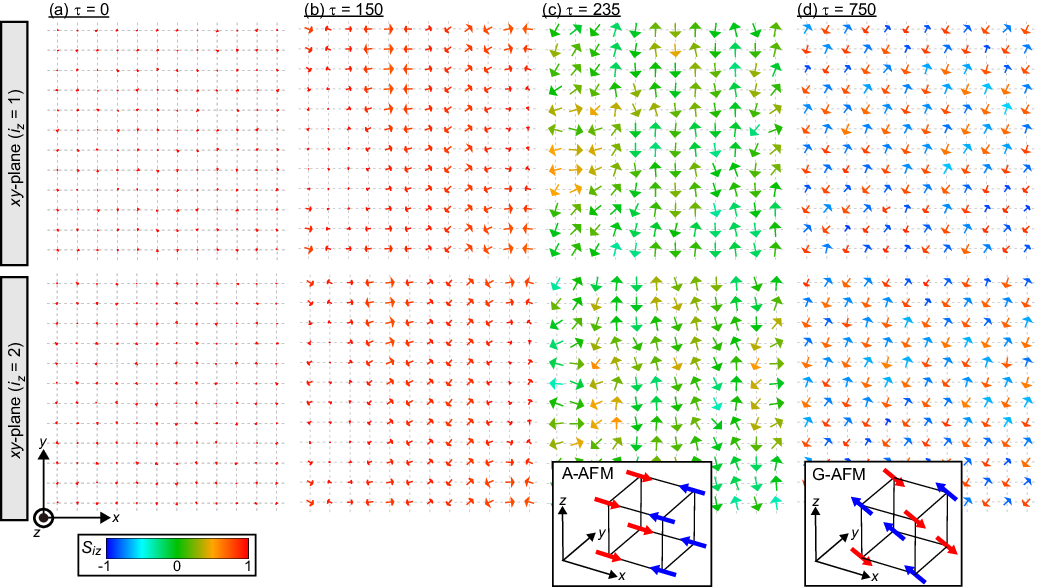}
\caption{Real-space spin configurations on the $xy$ planes at selected moments, i.e., (a) $\tau \leq 0$ (before irradiation), (b) $\tau$=150, (c) $\tau$=235, and (d) $\tau$=750, during the photoinduced magnetic phase transition from FM to G-AFM states under light irradiation with $\bm E$$\parallel$[100]. In (b)-(d), the arrows represent the in-plane spin components $(S_{ix}, S_{iy})$, while the colors represent the out-of-plane spin components $S_{iz}$. The simulations are performed for a system size of $N=12^3$ and the light field $\bm A(\tau)=A(\sin(\Omega\tau), 0, 0)$ with $A=2.0$ and $\Omega=0.5$.}
\label{Fig04}
\end{figure*}
The continued photoirradiation further reduces the bandwidth, and the excited electrons begin to occupy all the eigenstates constituting the contracted lower band. After $\tau=300$, the electron occupation $n_\mu$ becomes almost uniform in the lower band. Importantly, the system with this uniformly occupied lower band can be regarded as a pseudo half-filling system, where a situation similar to the half-filling case at equilibrium is realized. Consequently, the G-AFM state, which is the ground-state spin configuration at half filling in the equilibrium system (see Fig.~\ref{Fig01}) is realized under the photoirradiation.

Figure~\ref{Fig04} shows real-space spin configurations on neighboring two $xy$ planes with $i_z$=1 and $i_z$=2 at selected moments during the process of the photoinduced FM to G-AFM phase transition, i.e., (a) $\tau \leq 0$, (b) $\tau$=150 (99 fs), (c) $\tau$=235 (155 fs), and (d) $\tau$=750 (495 fs), where the in-plane components ($S_{ix}$, $S_{iy}$) are represented by arrows, while the out-of-plane components $S_{iz}$ are represented by colors. 
The parallel spin configurations along the $z$ axis in the initial FM state [Fig.~\ref{Fig04}(a)] gradually changes upon the photoirradiation with $\bm E$$\parallel$[100]. The spins start rotating within the $xy$ planes towards the head-to-head or tail-to-tail spin alignments, and the A-AFM correlation with staggered spin components along the $x$ axis is increased [Fig.~\ref{Fig04}(b)]. Comparing the two $xy$ planes with $i_z$=1 and $i_z$=2 in Fig.~\ref{Fig04}(b), we find that the spins are stacked ferromagnetically along the $z$ axis at this moment. Subsequently, many domains are nucleated in which the short-range A-AFM spin configuration is realized in the $z$-polarized FM background [Fig.~\ref{Fig04}(c)]. At this moment, the inter-layer spin correlation is still ferromagnetic. By further irradiation with light, the antiferromagnetic correlations grow also in the $y$ and $z$ directions to form a long-range G-AFM order in the final state [Fig.~\ref{Fig04}(d)]. 

\begin{figure}[tb]
\includegraphics[scale=0.5]{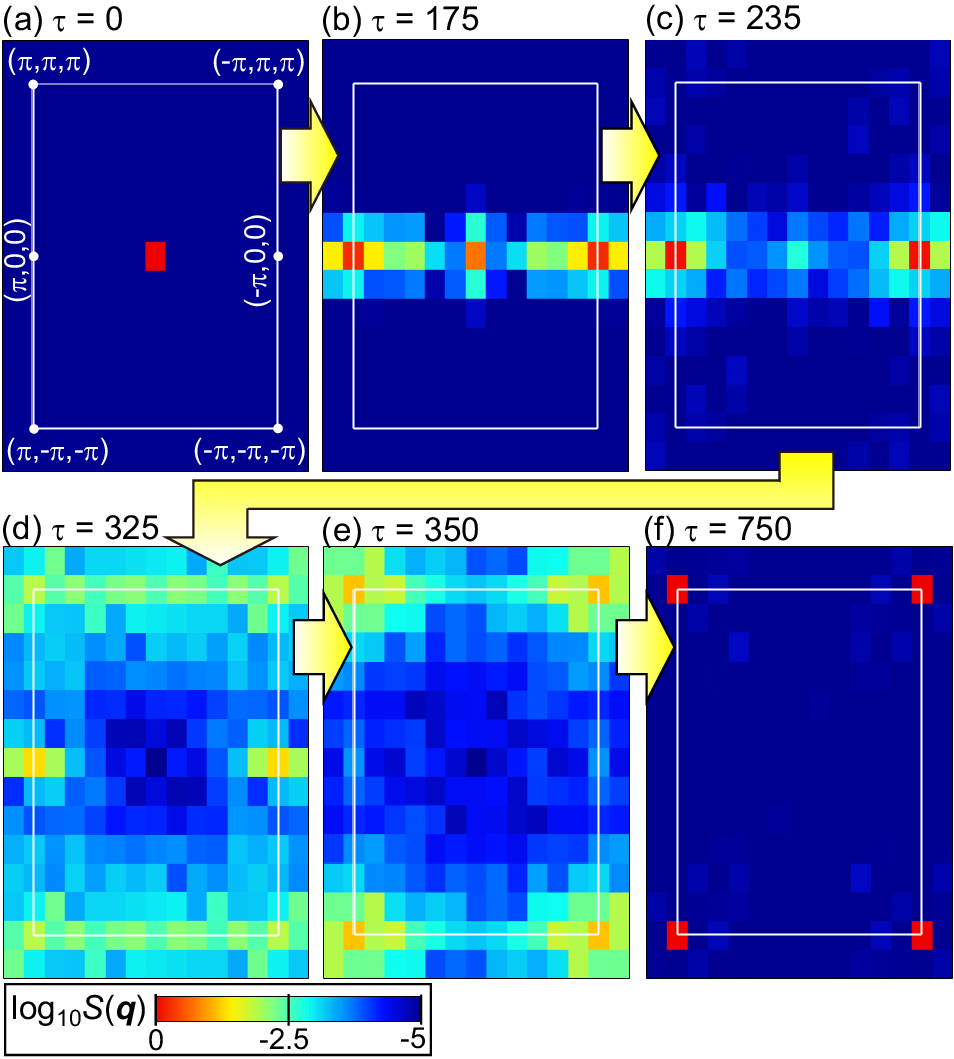}
\caption{Snapshots of the spin structure factor $S(\bm q)$ in the momentum space at selected moments, (a) $\tau \leq 0$ (before irradiation), (b) $\tau$=175, (c) $\tau$=235, (d) $\tau$=325, (e) $\tau$=350, and (f) $\tau$=750, during the photoinduced FM to G-AFM phase transition under light irradiation with $\bm E$$\parallel$[100]. The simulations are performed for a system size of $N=12^3$ and the light field $\bm A(\tau)=A(\sin(\Omega\tau), 0, 0)$ with $A=2.0$ and $\Omega=0.5$.}
\label{Fig05}
\end{figure}
The evolution of spin configuration can also be seen in snapshots of the spin structure factor $S(\bm q)$ in the momentum space. Figure~\ref{Fig05} shows those at selected moments, i.e, (a) $\tau \leq 0$ (before irradiation), (b) $\tau$=175, (c) $\tau$=235, (d) $\tau$=325, (e) $\tau$=350, and (f) $\tau$=750. For the initial state, $S(\bm q)$ has a large peak at $\bm q=(0,0,0)$ [Fig.~\ref{Fig05}(a)]. The photoirradiation first induces the increase of new peaks at $\bm q=(\pm\pi,0,0)$ and the decrease of the peak at $\bm q=(0,0,0)$ [Figs.~\ref{Fig05}(b) and (c)], indicating a growth of the A-AFM correlation and a suppression of the FM correlation. The further photoirradiation induces subsequent broad structures in $S(\bm q)$ along a line connecting $\bm q=(\pi,\pi,\pi)$ and $\bm q=(-\pi,\pi,\pi)$ and a line connecting $\bm q=(\pi,-\pi,-\pi)$ and $\bm q=(-\pi,-\pi,-\pi)$ [Fig.~\ref{Fig05}(d)]. These structures evolve to form peaks at these four momentum points [Figs.~\ref{Fig05}(e) and (f)], indicating the emergence of G-AFM phase under the photoirradiation.

\begin{figure}[tb]
\includegraphics[scale=1.0]{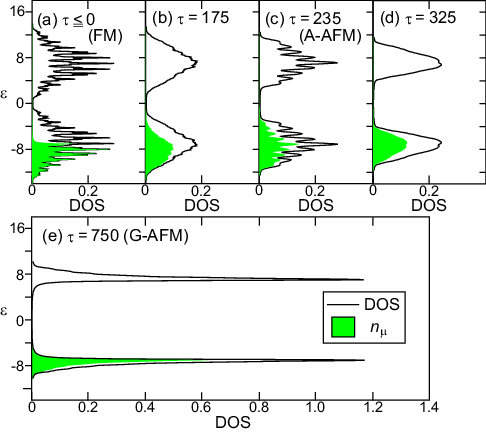}
\caption{Snapshots of the time-varying density of states and the electron occupation $n_\mu$ as functions of the energy $\varepsilon$ at selected moments, i.e., (a) $\tau \leq 0$ (before irradiation), (b) $\tau$=175, (c) $\tau$=235, (d) $\tau$=325, and (e) $\tau$=750, during the photoinduced FM to G-AFM phase transition under light irradiation with $\bm E$$\parallel$[100] for a system size of $N=12^3$ and the light field $\bm A(\tau)=A(\sin(\Omega\tau), 0, 0)$ with $A=2.0$ and $\Omega=0.5$.}
\label{Fig06}
\end{figure}
Figure~\ref{Fig06} shows snapshots of the density of states and the electron occupation as functions of the energy $\varepsilon$ at selected moments, i.e., (a) $\tau \leq 0$ (before irradiation), (b) $\tau$=175, (c) $\tau$=235, (d) $\tau$=325, and (f) $\tau$=750. Before the photoirradiation, the lower half of the lower band separated from the upper band by the exchange gap is realized for the present quarter filling system with $n_{\rm e}=0.25$ [Fig.~\ref{Fig06}(a)]. Noticeably the photoirradiation causes the contraction of bandwidth and the increase of energy gap between the upper and lower bands. The electrons are excited from the lower to higher lying states within the lower band. However, at the moments when the A-AFM correlation is grown [Figs.~\ref{Fig06}(b) and (c)], the upper portion of the moderately contracted lower band remains to be unoccupied, which is far from the pseudo half-filling condition. On the contrary, at the moments when the G-AFM correlation is grown, even the upper portion of the lower band is occupied [Figs.~\ref{Fig06}(d) and (e)]. In particular, when the G-AFM long range order is realized, the electron occupation of the upper portion is significantly enhanced [Fig.~\ref{Fig06}(e)], because of which the system behaves like a half-filling system at equilibrium.

\begin{figure}[tb]
\includegraphics[scale=1.0]{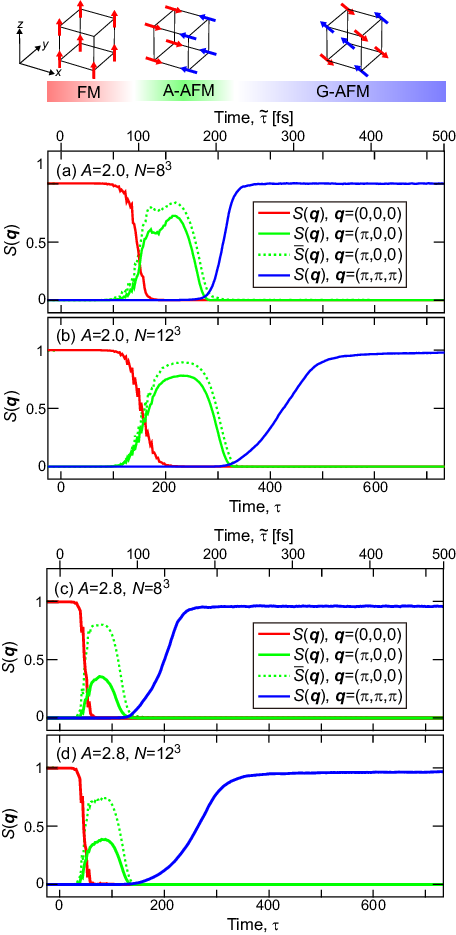}
\caption{Time profiles of the spin structure factors $S(\bm q)$ at $\bm q=(0,0,0)$, $\bm q=(\pi,0,0)$ and $\bm q=(\pi,\pi,\pi)$ for the photoinduced FM to G-AFM phase transition via the phase with A-AFM spin correlations under light irradiation with $\bm E$$\parallel$[100] for different light-field intensities $A$ and system sizes $N$, i.e., (a) $A=2.0$ and $N=8^3$, (b) $A=2.0$ and $N=12^3$, (c) $A=2.8$ and $N=8^3$, and (d) $A=2.8$ and $N=12^3$. Time profiles of the integrated spin structure factor $\bar{S}(\bm q)$ for $\bm q=(\pi,0,0)$ are also plotted by dashed lines. Here the upper horizontal axes represent the real time $\tilde{\tau}$ in the units of fs when $t$=1 eV.}
\label{Fig07}
\end{figure}
In order to discuss dependence on the system size and the light-field intensity, we also perform numerical simulations for different two system sizes, i.e., $N=8^3$ and $N=12^3$ and different two light-field intensities, i.e., $A=2.0$ and $A=2.8$. For all the four combinations, we again observe the photoinduced magnetic phase transition from FM to G-AFM phases via a transient phase in which the A-AFM correlation increases under irradiation with linearly polarized light $\bm E$$\parallel$[100]. Figure~\ref{Fig07} shows calculated time profiles of $S(\bm q)$ at $\bm q=(0,0,0)$, $\bm q=(\pi,\pi,0)$ and $\bm q=(\pi,\pi,\pi)$ for (a) $A=1.6$ and $N=8^3$, (b) $A=1.6$ and $N=12^3$, (c) $A=2.0$ and $N=8^3$, and (d) $A=2.0$ and $N=12^3$. Comparison between Figs.~\ref{Fig07}(a) and (b) and that between Figs.~\ref{Fig07}(c) and (d) reveal that the A-AFM correlation quantified by $S(\bm q)$ at $\bm q=(\pi,0,0)$ does not depend on the system size $N$ so much. On the contrary, it sensitively depends on the light-field intensity $A$ as seen in comparison between Figs.~\ref{Fig07}(a) and (c) and that between Figs.~\ref{Fig07}(b) and (d). The latter result indicates that for a stronger laser light, the A-AFM correlation in the transient phase becomes shorter ranged with dynamically nucleated smaller sized A-AFM domains.

To confirm this aspect, we also calculate the integrated spin structure factor $\bar{S}(\bm q)$, which is defined by,
\begin{align}
\label{eq:IntSq}
\bar{S}(\bm q,\tau) \equiv \sum_{i=1}^{27} S(\bm q_i,\tau),
\end{align}
where $\bm q_i=(q_{ix}, q_{iy}, q_{iz})$ represents ($3^3-1$) momentum points around $\bm q$ and $\bm q$ itself. For $\bm q=(\pi,0,0)$, $q_{ix}=\pi, \pi \pm 2\pi/N_x$, $q_{iy}=0, \pm 2\pi/N_y$ and $q_{iz}=0, \pm 2\pi/N_z$. In Figs.~\ref{Fig07}(a)-(d), calculated time profiles of $\bar{S}(\bm q)$ for a peak around $\bm q=(\pi,0,0)$ are plotted by dashed lines, which quantifies a growth of the short-range A-AFM spin correlations. We find that the difference between $S(\bm q)$ and $\bar{S}(\bm q)$ at $\bm q=(\pi,0,0)$ is smaller for a weaker light field of $A=2.0$ as seen in Figs.~\ref{Fig07}(a) and (b). On the contrary,  Figs.~\ref{Fig07}(c) and (d) show that for a stronger light field of $A=2.8$, the difference is pronounced, indicating that the peak around $\bm q$ becomes broader. This means that the dynamically nucleated A-AFM domains tend to be smaller in size and short-ranged for a stronger light field. Note that we have investigated cases with various strengths of exchange coupling $J$ and found that the photoinduced phase transition from the FM to the G-AFM phase predicted in this study is robust and occurs over a broad range of $J$. Further details of these findings are provided in the Appendix.~\ref{sec:Appendix1}.

\subsection{Linearly polarized light with $\bm E$$\parallel$[110]}
\begin{figure*}[tb]
\includegraphics[scale=1.0]{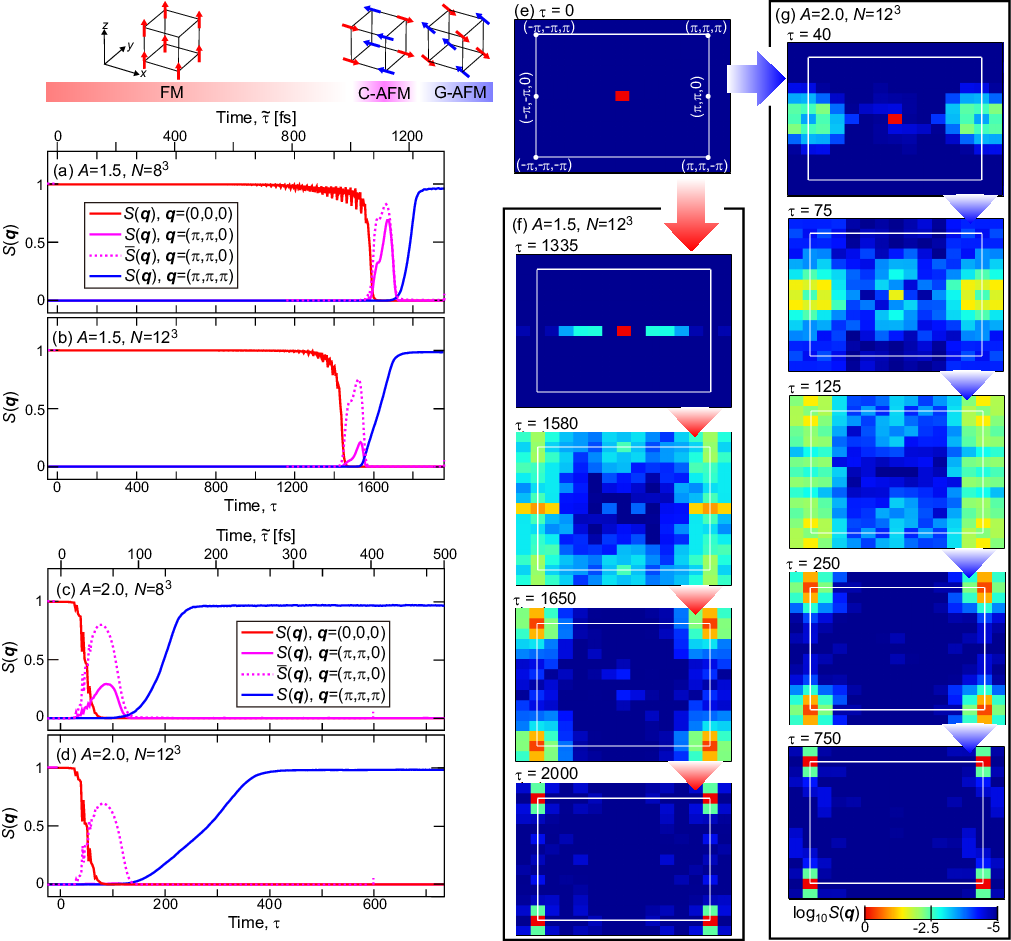}
\caption{Results for the linearly polarized light with $\bm E$$\parallel$[110], which indicate a photoinduced magnetic phase transition from FM to G-AFM phases via a transient phase with C-AFM spin correlations. The simulations are performed for the light field $\bm A(\tau)=A(\sin(\Omega\tau), \sin(\Omega\tau), 0)$ with $\Omega=0.5$. (a)-(d) Time profiles of the spin structure factors $S(\bm q)$ at $\bm q=(0,0,0)$, $\bm q=(\pi,\pi,0)$ and $\bm q=(\pi,\pi,\pi)$, which quantify the FM, C-AFM and G-AFM spin correlations, respectively, for different light-field intensities $A$ and system sizes $N$, i.e., (a) $A=1.5$ and $N=8^3$, (b) $A=1.5$ and $N=12^3$, (c) $A=2.0$ and $N=8^3$, and (d) $A=2.0$ and $N=12^3$. Time profiles of the integrated spin structure factor $\bar{S}(\bm q)$ for $\bm q=(\pi,\pi,0)$ are also plotted by dashed lines.
(e) Spin structure factor $S(\bm q)$ in the momentum space for the initial FM state. (f),~(g) Snapshots of $S(\bm q)$ at selected moments for different light-field intensities, (f) $A=1.5$ and (g) $A=2.0$, where the system size is $N=12^3$ for both cases.}
\label{Fig08}
\end{figure*}
We next discuss the results for the linearly polarized light with $\bm E$$\parallel$[110]. The simulations are performed for the light field $\bm A(\tau)=A(\sin(\Omega\tau), \sin(\Omega\tau), 0)$ with $\Omega=0.5$. Figures~\ref{Fig08}(a)-(d) show calculated time profiles of the spin structure factors $S(\bm q)$ at $\bm q=(0,0,0)$, $\bm q=(\pi,\pi,0)$, and $\bm q=(\pi,\pi,\pi)$, which quantify the FM, C-AFM and G-AFM spin correlations, respectively, for different light-field intensities $A$ and system sizes $N$, i.e., (a) $A=1.5$ and $N=8^3$, (b) $A=1.5$ and $N=12^3$, (c) $A=2.0$ and $N=8^3$, and (d) $A=2.0$ and $N=12^3$. Here the C-AFM is an antiferromagnetic state in which the localized spins are aligned antiferromagnetically along the $x$ and $y$ axes and ferromagnetically along the $z$ axis. All these figures indicate that a photoinduced magnetic phase transition from FM to G-AFM phases occurs via a transient phase with C-AFM spin correlation. 

Noticeably, the type of spin correlation in the transient phase is distinct from the above-discussed case with $\bm E$$\parallel$[100]. Specifically, the A-AFM spin correlation is observed in the former case with $\bm E$$\parallel$[100], where staggered spin alignments are realized only in the $x$ direction. On the contrary, the C-AFM spin correlation is observed in the present case with $\bm E$$\parallel$[110], where the staggered spin alignments are realized both in the $x$ and $y$ directions. Note that the interlayer spin correlation along the $z$ axis is ferromagnetic for both cases. This difference can be attributed to the difference in the light-induced band-structure modulation due to the dynamical localization effect. For $\bm E$$\parallel$[100], only the transfer integrals on bonds along the $x$ axis is effectively suppressed, while those on bonds along the $x$ and $y$ axes are suppressed for $\bm E$$\parallel$[110]. The photoinduced band contraction and resulting enhancement of the Fermi-surface nesting can occur both in the $k_x$ and $k_y$ directions for the present case with $\bm E$$\parallel$[110]. This situation favors the C-AFM spin correlation with antiferromagnetic coupling in both $x$ and $y$ directions. 

The dependence on the system size $N$ can be discussed again by comparing Figs.~\ref{Fig08}(a) and (b) as well as Figs.~\ref{Fig08}(c) and (d). We can also discuss the dependence on the light-field intensity $A$ by comparing Figs.~\ref{Fig08}(a) and (c) as well as Figs.~\ref{Fig08}(b) and (d). It is found that the C-AFM correlation quantified by $S(\bm q)$ at $\bm q=(\pi,\pi,0)$ in the transient phase strongly depends on the system size $N$, which significantly decreases as the system size increases from $N=8^3$ to $N=12^3$. This behavior is in sharp contrast to the A-AFM correlation in the transient phase under $\bm E$$\parallel$[100], which does not show any clear difference in amplitude of $S(\bm q)$ at $\bm q=(\pi,0,0)$ between the cases with $N=8^3$ and $N=12^3$. 

The decrease of the transient C-AFM correlation with increasing system size might indicate that the C-AFM domains are dynamically nucleated under light irradiation and the typical size of the nucleated domains does not change so much irrespective of the system size. If the typical domain size is $L \times L$ sites with $L \sim 4-6$ in the $xy$ planes and the domains are nucleated rather randomly in space and time, the peaks in $S(\bm q)$ at $\bm q=(\pi,\pi,0)$ should be more diffusive and broader as the system size increases from $N=8^3$ to $N=12^3$. Indeed, the integrated spin structure factors $\bar{S}(\bm q)$ for $\bm q=(\pi,\pi,0)$ has almost the same intensities for all the cases, indicating that the short-range C-AFM correlations are grown in the transient phase irrespective of the system size and the light-field intensity.

In Figs.~\ref{Fig08}(e)-(g), we show snapshots of the spin structure factor $S(\bm q)$ in the momentum space at selected moments. More specifically, Fig.~\ref{Fig08}(e) shows a snapshot for the initial FM state ($\tau \leq 0$), while Figs.~\ref{Fig08}(f) and (g) show calculated time evolutions for  cases with different light intensities, i.e., (f) $A=2.1$ and $N=12^3$, and (g) $A=2.8$ and $N=12^3$. Noticeably, peaks of $S(\bm q)$ in the transient phase are very broad and diffusive, which have significant weight at the points away from the momentum $\bm q=(\pi,\pi,0)$ corresponds to the C-AFM correlation. Moreover, the peaks are broader and more diffusive for a stronger light field. This would support the validity of the above argument. We also show snapshots of the spatial spin configurations on a $xy$ plane ($i_z=1$) for systems with different sizes, i.e., $N=8^3$ to $N=12^3$, at the moments when $S(\bm q)$ at $\bm q=(\pi,\pi,0)$ takes almost maximum. We find that typical sizes of the nucleated C-AFM domains do not differ so much between the two cases, which also supports the above argument.

\begin{figure}[tb]
\includegraphics[scale=1.0]{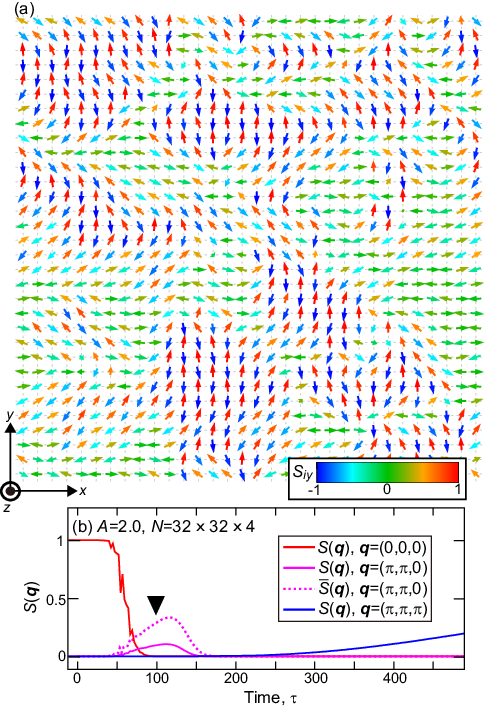}
\caption{Snapshot of the spin configuration on a selected $xy$ plane for a larger system size of $N=32 \times 32 \times 4$ at a moment when the spin structure factor $S(\bm q)$ at $\bm q=(\pi,\pi,0)$, which quantifies the C-AFM spin correlation, takes nearly a maximum value, i.e., at $\tau=100$. The spin vectors $\{\bm S_i\}$ have dominant in-plane components $(S_{ix},S_{iy})$, which are represented by arrows. The colors represent the $S_{iy}$ component. (b) Time profiles of the spin structure factors $S(\bm q)$ and $\bar{S}(\bm q)$. The time at which the snapshot is taken is indicated by the inverted triangle. The simulations are performed for the light field $\bm A(\tau)=A(\sin(\Omega\tau), \sin(\Omega\tau), 0)$ with $A=2.0$.}
\label{Fig09}
\end{figure}
In addition, we examine the transient phase with C-AFM spin correlation in a larger-sized system of $N=32 \times 32 \times 4$. Figure~\ref{Fig09} shows a snapshot of the spin configuration on a selected $xy$ plane ($i_z=1$) at a moment when the spin structure factor $S(\bm q)$ at $\bm q=(\pi,\pi,0)$, which quantifies the C-AFM spin correlation, takes a maximum. The spin vectors $\{\bm S_i\}$ have dominant in-plane components $(S_{ix},S_{iy})$, which are represented by arrows, whereas their out-of-plane components $S_{iz}$ are very small. The colors represent the $S_{iy}$ component. Noticeably, there appear many small C-AFM domains with staggered spin alignments along both $x$ and $y$ axes. 

The staggered spins are oriented in various directions depending on the domains. Specifically, areas with red and blue arrows are domains with spins oriented in the $\pm x$ direction, while those with green arrows are domains with spins oriented in the $\pm y$ direction. It is also found that the typical size of these C-AFM domains in the transient phase does not change so much even if we vary the system size according to a comparison between this snapshot and that for $N=12^3$ (not shown). These two facts might be attributed to the diffusive peaks in $S(\bm q)$ at $\bm q=(\pi,\pi,0)$ and the suppression of peak heights with increasing system size. It also indicates that the apparent suppression of $S(\bm q)$ at $\bm q=(\pi,\pi,0)$ with increasing system size does not necessarily mean that the C-AFM spin correlation is weaker in a larger system. As seen in Fig.~\ref{Fig09}(a), the $xy$ plane in the larger system of $N=32 \times 32 \times 4$ are fully covered by the C-AFM domains, although $S(\bm q)$ at $\bm q=(\pi,\pi,0)$ is much smaller than unity.

\subsection{Circularly polarized light}
\begin{figure*}[tb]
\includegraphics[scale=1.0]{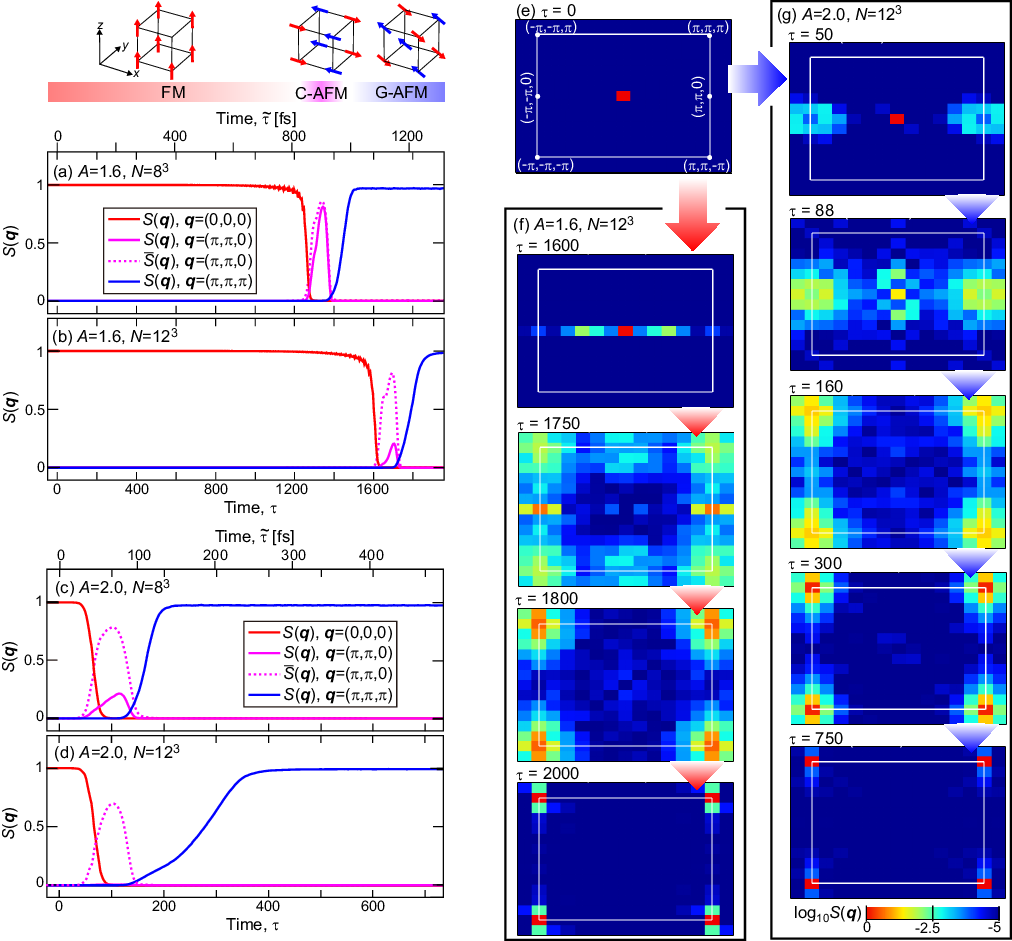}
\caption{Results for the circularly polarized light, which again indicate a photoinduced magnetic phase transition from FM to G-AFM phases via a transient phase with C-AFM spin correlations. The simulations are performed for the light field $\bm A(\tau)=A(\cos(\Omega\tau), \sin(\Omega\tau), 0)$ with $\Omega=0.5$. (a)-(d) Time profiles of the spin structure factors $S(\bm q)$ at $\bm q=(0,0,0)$, $\bm q=(\pi,\pi,0)$ and $\bm q=(\pi,\pi,\pi)$, which quantify the FM, C-AFM and G-AFM spin correlations, respectively, for different light-field intensities $A$ and system sizes $N$, i.e., (a) $A=1.6$ and $N=8^3$, (b) $A=1.6$ and $N=12^3$, (c) $A=2.0$ and $N=8^3$, and (d) $A=2.0$ and $N=12^3$. Time profile of the integrated spin structure factor $\bar{S}(\bm q)$ for $\bm q=(\pi,\pi,0)$ are also plotted by dashed lines.
(e) Spin structure factor $S(\bm q)$ in the momentum space for the initial FM state. (f),~(g) Snapshots of $S(\bm q)$ at selected moments for different light-field intensities, (f) $A=1.6$ and (g) $A=2.0$, where the system size is $N=12^3$ for both cases.}
\label{Fig10}
\end{figure*}
We finally show results for the circularly polarized light. The simulations are performed for the light field $\bm A(\tau)=A(\sin(\Omega\tau), \cos(\Omega\tau), 0)$ with $\Omega=0.5$. Figures~\ref{Fig10}(a)-(d) show time profiles of the spin structure factors $S(\bm q)$ at $\bm q=(0,0,0)$, $\bm q=(\pi,\pi,0)$ and $\bm q=(\pi,\pi,\pi)$ for different light-field intensities $A$ and system sizes $N$, i.e., (a) $A=1.6$ and $N=8^3$, (b) $A=1.6$ and $N=12^3$, (c) $A=2.0$ and $N=8^3$, and (d) $A=2.0$ and $N=12^3$. Again we observe a photoinduced magnetic phase transition from FM to G-AFM phases via a transient phase with C-AFM spin correlation, similar to the case with linearly polarized light with $\bm E$$\parallel$[110].

The emergence of the C-AFM spin correlation in the transient phase can be explained by the same argument as the case of the linearly polarized light with $\bm E$$\parallel$[110]. More specifically, the circularly polarized light should effectively reduce the transfer integrals on bonds along the $x$ and $y$ axes via the dynamical localization effect, and, therefore, the contraction of the band structure and the enhancement of the Fermi-surface nesting would occur in similar manners as those for the light with $\bm E$$\parallel$[110]. The dependence of the C-AFM correlation on the system size $N$ and the light-field intensity $A$ are also similar to the case with $\bm E$$\parallel$[110]. Namely, the spin structure factor $S(\bm q)$ at $\bm q=(\pi,\pi,0)$ decreases as either $N$ or $A$ increases. These behaviors can be explained by the same argument. 

\begin{figure*}[tb]
\includegraphics[scale=1.0]{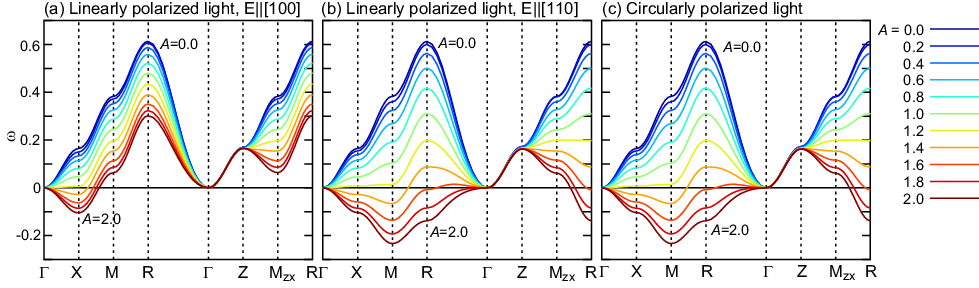}
\caption{(a) Calculated magnon dispersion relations for various light intensities under application of linearly polarized light $\bm E$$\parallel$[100] with $\bm A(\tau)=A(\sin(\Omega\tau),0,0)$. (b) Those for linearly polarized light $\bm E$$\parallel$[110] with $\bm A(\tau)=A(\sin(\Omega\tau),\sin(\Omega\tau),0)$. (c) Those for circularly polarized light with $\bm A(\tau)=A(\cos(\Omega\tau),\sin(\Omega\tau),0)$. The light frequency is fixed at $\Omega$=0.5.}
\label{Fig11}
\end{figure*}
Emergence of the transient phase with a low-dimensional antiferromagnetic correlation and its light-polarization dependence can be argued as instabilities of the ferromagnetic ground state towards the A-AFM or C-AFM state under photoirradiation. For this purpose, we calculate the magnon dispersion relation in the photodriven system for various cases with different light polarizations using a theoretical formalism based on the Floquet Green's functions~\cite{Ono2018}. With this formalism, we can calculate magnon spectra in nonequilibrium steady states realized by light through effectively mapping problems of the time-periodically driven system onto problems of a certain equilibrium system (see appendix~\ref{sec:Appendix2}). 

Figure~\ref{Fig11} shows calculated magnon dispersion relations for various light intensities under the light irradiation with different light polarizations, that is, (a) linearly polarized light $\bm E$$\parallel$[100] with $\bm A(\tau)=A(\sin(\Omega\tau),0,0)$, (b) linearly polarized light $\bm E$$\parallel$[110] with $\bm A(\tau)=A(\sin(\Omega\tau),\sin(\Omega\tau),0)$, and (c) circularly polarized light with $\bm A(\tau)=A(\cos(\Omega\tau),\sin(\Omega\tau),0)$. Here the light intensity and frequency are fixed at $A$=2.0 and $\Omega$=0.5. 

In Fig.~\ref{Fig11}(a) for $\bm E$$\parallel$[100], the magnon dispersion relations exhibit a clear softening with increasing light intensity $A$ at X point $\bm q=(\pi,0,0)$. As the light intensity $A$ increases, the magnon energy touches zero when $A \approx 1.2$ at this momentum point and becomes negative with further increasing $A$. This indicates that the FM ground state is destabilized by the intense light with $\bm E$$\parallel$[100] and an instability towards the A-AFM state emerges. Importantly, the magnon energy becomes negative only at X point in the case of $\bm E$$\parallel$[100]. The emergence of clear A-AFM correlation observed in the simulations can be attributed to this aspect. 

On the contrary, in Fig.~\ref{Fig11}(b) for $\bm E$$\parallel$[110], the softening occurs at different momentum points in multiple stages. As $A$ increases, the magnon energy first touches zero when $A \approx 1.2$ both at X point  $\bm q=(\pi,0,0)$ and at M point  $\bm q=(\pi,\pi,0)$. The energy becomes negative with further increasing $A$ at these points, whereas the energy is still positive at R point $\bm q=(\pi,\pi,\pi)$. As $A$ is further increased, the magnon band touches the zero energy also at R point when $A \approx 1.6$. Subsequently, the energy at R point becomes negative with further increasing $A$. 

These aspects indicate that destabilization of the FM ground state is also caused by the intense light with $\bm E$$\parallel$[110] when $A \gtrsim 1.2$, but the type of spin correlation in the transient phase varies below and above $A \approx 1.6$. Specifically, for $1.2 \lesssim A \lesssim 1.6$, the magnon band energy is negative at X and M points but positive at R point, indicating that the AFM correlation appears within the $xy$ planes, while the FM correlation still survives along the $z$ axis, which results in the emergence of the pronounced C-AFM correlation in the transient phase as observed in Fig.~\ref{Fig09}(b). On the other hand, for $A \gtrsim 1.6$, the magnon energy becomes negative at R point as well as X and M points, indicating that the spin correlation along the $z$ axis changes from FM to AFM. In this case, the C-AFM correlation becomes weak in the transient phase as observed in Fig.~\ref{Fig09}(d). Instead, the transient phase can be regarded as a state in which the long-range AFM correlation is grown within the planes, but the spin correlation is still short ranged along the stacking direction.

This means that the nature of emergent spin correlation in the transient phase varies depending on the light intensity when the system is irradiated with linearly polarized light $\bm E$$\parallel$[110]. This interesting phenomenon can be expected for the system irradiated with circularly polarized light, in which the softening of magnon band dispersion and the instability of FM ground state occur in a similar manner to the case of linearly polarized light $\bm E$$\parallel$[110] as seen in Fig.~\ref{Fig11}(c). The instability with negative magnon energy is observed in a broad area of the momentum space including X, M and R points. In contrast, such light-intensity dependence of  the transient spin correlation should be difficult for the system irradiated with linearly polarized light $\bm E$$\parallel$[100], because the instability with negative magnon energy occurs always at X point corresponding to the A-AFM spin correlation.

\section{Conclusion}
We have theoretically studied photoinduced magnetic phase transitions and their dynamical processes in the Kondo-lattice model on a cubic lattice. First, by numerically solving the coupled time-evolution equations for the electron system and the localized-spin system, we have demonstrated that the photoinduced phase transition from the ground-state ferromagnetism to a three-dimensional antiferromagnetism, called G-type antiferromagnetism, occurs in the photodriven system. By examining the time evolutions of energy band structure and electron occupation under the light irradiation, we have found that a narrowing of the bandwidth and an increase of the exchange gap due to dynamical localization effects occur. Furthermore, we have revealed that all the electron states constituting the lower band separated from the upper band by the exchange gap become partially but almost uniformly occupied as a consequence of the photoexcitation and relaxation of electrons. This pseudo half-filling electronic structure stabilizes the G-type antiferromagnetism, which is realized in the half-filling ground state of the equilibrium system. 

By studying time evolutions of the spin structure factor and the spatial configuration of the localized spins, we have found that lower-dimensional antiferromagnetic correlations of A-type and C-type appear in the transient phase of this dynamical phase transition. The type of antiferromagnetic correlation in the transient phase strongly depends on the polarization of irradiated light. Specifically, the A-type antiferromagnetic correlation appears for linearly polarized light with $\bm E$$\parallel$[100], while the C-type appears for linearly polarized light with $\bm E$$\parallel$[110] and circularly polarized light. This light-polarization dependence is attributable to the effective decrease of the transfer integrals on bonds along the polarization direction and the resulting bandwidth narrowing in the corresponding momentum direction, which leads to an enhancement of the Fermi-surface nesting and the increase of the antiferromagnetic correlation in this direction. We have also calculated the magnon spectra of the photodriven system using a theory based on the Floquet Green's function formalism and have revealed that the instability to the A-type or C-type antiferromagnetic state is caused by a softening of the magnon band dispersion at a specific momentum point, which depends on the light polarization. Our theoretical findings provide important insights into the understanding of photoinduced magnetic phase transitions in three-dimensional spin-charge coupled magnets.

\section{Appendix: Dependence on the Kondo-coupling strength}
\label{sec:Appendix1}
\begin{figure}[tb]
\includegraphics[scale=1.0]{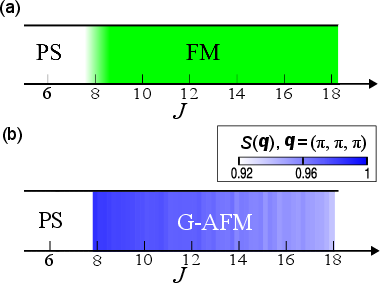}
\caption{(a) Ground-state phase diagram as a function of the Kondo-coupling constant $J$ at the electron filling of $n_{\rm e}=1/4$. Here PS denotes the phase separation. (b) Phase diagram for nonequilibrium steady states under photoirradiation as a function of $J$ at $n_{\rm e}=1/4$. Calculated values of the spin structure factor $S(\pi,\pi,\pi)$ at $\tau=2000$ are presented by colors. The system is irradiated by linearly polarized light $\bm{A}(\tau)=A\sin\Omega\tau(1,0,0)$ with $A=2.0$ and $\Omega=0.5t$.}
\label{Fig12}
\end{figure}
We have demonstrated that the photoinduced magnetic phase transition from the FM to G-AFM phase occurs in the Kondo-lattice model at $J=14t$. Here, we show that this phenomenon is not limited to this specific value of the Kondo-coupling constant but can be observed over a broad range of $J$ values, provided that the ground state is ferromagnetic. For example, when the electron filling is $n_{\rm e}=1/4$, the ferromagnetic ground state appears when $J/t \gtrapprox 8$ [Fig.~\ref{Fig12}(a)]. To examine the robustness of the photoinduced phase transition to the G-AFM phase, we performed simulations for various values of $J$. Figure~\ref{Fig12}(b) presents the obtained phase diagram for nonequilibrium steady states under photoirradiation as a function of $J$ at $n_{\rm e}=1/4$. The calculated values of the spin structure factor $S(\pi,\pi,\pi)$, which characterizes the G-AFM order, after a sufficient duration of photoirradiation (at $\tau=2000$) are shown in color. Here, the system is irradiated by linearly polarized light $\bm{A}(\tau)=A\sin\Omega\tau(1,0,0)$ with $A=2.0$ and $\Omega=0.5t$. The phase diagram indicates that the photoinduced phase transition to the G-AFM phase can occur in the cubic Kondo-lattice model for a wide range of $J$ values, including relatively small values around $J \approx 8t$.

\section{Appendix: Floquet Green's function formalism}
\label{sec:Appendix2}
In this appendix, we explain the Floquet Green's function formalism for calculating the magnon dispersion relation in the photodriven system.

\subsection{Floquet theory}
The Floquet theory is a theoretical scheme to solve a problem of the time-periodically driven system by mapping the system onto an effective static problem. The theory can be applied to an ordinary differential equation,
\begin{align}
\label{eq:Flqt1}
\frac{d}{d\tau}x(\tau)=C(\tau)x(\tau),
\end{align}
where $C(\tau)$ is a periodic function with respect to $\tau$. We start with the following time-dependent Schr\"odinger equation,
\begin{align}
\label{eq:tScheq}
i\frac{d}{d\tau}\Psi(\tau)=\mathcal{H}(\tau)\Psi(\tau),
\end{align}
where the Hamiltonian $\mathcal{H}(\tau)$ is time-periodic with a periodicity of $T_{\rm p}$. In general, the solution of this equation is given in the form,
\begin{align}
\label{eq:Flqt2}
\Psi_\alpha(\tau)=e^{-i\varepsilon_\alpha \tau}u_\alpha(\tau).
\end{align}
Here $\varepsilon_\alpha$ is a real quantity called quasi energy, and $u_\alpha(\tau)$ is a time-periodic function with a periodicity of $T_{\rm p}$. We introduce the Fourier transformations for the time-periodic functions $\mathcal{H}(\tau)$ and $u_\alpha(\tau)$ as,
\begin{align}
\label{eq:FTH}
\mathcal{H}(\tau)=\sum_{n=-\infty}^{\infty} e^{-in\Omega \tau} \mathcal{H}^{(n)},
\\
\label{eq:FTu}
u_\alpha(\tau)=\sum_{n=-\infty}^{\infty} e^{-in\Omega \tau}u_\alpha^{(n)},
\end{align}
with $T_{\rm p}=2\pi/\Omega$.
By substituting Eqs.~\eqref{eq:FTH} and \eqref{eq:FTu} into the time-dependent Schr\"odinger equation in Eq.~\eqref{eq:tScheq}, we obtain the following eigenequation,
\begin{align}
\label{eq:Fegeq}
\sum_{n=-\infty}^{\infty} (\mathcal{H}_{mn}-n\delta_{mn}\Omega)u_\alpha^{(n)}
=\varepsilon_\alpha u_\alpha^{(m)}
\end{align}
where
\begin{align}
\label{eq:Hmn}
\mathcal{H}_{mn} \equiv \mathcal{H}^{(m-n)}.
\end{align}
In this way, the time-dependent Schr\"odinger equation in Eq.~\eqref{eq:tScheq} is mapped on the infinite-dimensional static eigenvalue problem.

In the photodriven system, the time-dependence of the Hamiltonian originates from time-periodic Peierls phases attached to the transfer integrals as,
\begin{align}
\label{eq:tdepH}
\mathcal{H}(\tau)=\sum_{i,j,\sigma}t_{ij}\exp[-i \bm A(\tau) \cdot (\bm r_i-\bm r_j)]
c^\dag_{i\sigma} c_{j\sigma},
\end{align}
where $\bm A(\tau)$ is the vector potential of light electromagnetic field. Here the effect of light irradiation can be taken into account by replacing the transfer integrals as,
\begin{align}
\label{eq:tdept}
t_{ij} \;\Rightarrow \;t_{ij}\exp[-i\bm A(\tau) \cdot (\bm r_i-\bm r_j)],
\end{align}
where we adopt the natural units $\hbar$=$e$=1. In the momentum representation, the Hamiltonian is given in the form,
\begin{align}
\label{eq:tdepHk}
\mathcal{H}(\tau)=\sum_{\bm k,\sigma}\varepsilon_{\bm k+\bm A(\tau)}
c^\dag_{\bm k\sigma} c_{\bm k\sigma}.
\end{align}
Namely, the effect of light irradiation is considered by replacing the momentum as,
\begin{align}
\label{eq:tdepk}
\bm k \;\Rightarrow \; \bm k+\bm A(\tau).
\end{align}

In the Fourier transformation of the time-dependent Hamiltonian $\mathcal{H}(\tau)$ for the Floquet formalism, we need to perform the Fourier transformation of $\varepsilon_{\bm k+\bm A(\tau)}$ as,
\begin{align}
\label{eq:FTeps}
\varepsilon_{\bm k}^{(n)}=\int_{-T_{\rm p}/2}^{T_{\rm p}/2}\;\frac{d\tau}{T_{\rm p}}
e^{in\Omega \tau}\varepsilon_{\bm k+\bm A(\tau)}.
\end{align}
Using the following formula
\begin{align}
\label{eq:Bfnc1}
&\int_{-T_{\rm p}/2}^{T_{\rm p}/2}\;\frac{d\tau}{T_{\rm p}}
\cos[\beta+\alpha\sin(\Omega \tau+\varphi)] e^{in\Omega \tau}
\notag\\ & \hspace{0.5cm}
=\frac{1}{2}J_n(\alpha) \left\{
(-1)^n e^{i\beta}+e^{-i\beta}\right\} e^{-in\varphi},
\end{align}
we calculate the integral in Eq.~\eqref{eq:FTeps} to obtain,
\begin{align}
\label{eq:Bfnc2}
\varepsilon_{\bm k}^{(n)}=
-t\sum_{\bm r=\hat{\bm x},\hat{\bm y},\hat{\bm z}} J_n(\bm A \cdot \bm r)
\left\{(-1)^n e^{i\bm k \cdot \bm r}+e^{-i\bm k \cdot \bm r}\right\} 
e^{-in\phi},
\end{align}
where $J_{n}(z)$ is the Bessel's function of the first kind, which is defined by,
\begin{align}
\label{eq:Bfnc3}
J_n(z)=\int_{-\pi}^\pi \frac{d\theta}{2\pi} e^{iz\sin\theta-in\theta}
=\int_{-\pi}^\pi \frac{d\theta}{2\pi} e^{in\theta-iz\sin\theta}.
\end{align}
Here we consider the transfer integrals $t$ only on the nearest neighbor bonds, and the following relation is used in the calculation,
\begin{align}
\label{eq:Bfnc4}
J_{-n}(z)=(-1)^nJ_n(z).
\end{align}

The phase $\phi$ differs depending on the light polarization. In the cases of linearly polarized light with $\bm A(\tau)=A\sin(\Omega \tau)(1,0,0)$ and $\bm A(\tau)=A\sin(\Omega \tau)(1,1,0)$, the phase $\phi$ is zero ($\phi=0$). On the contrary, in the case of circularly polarized light with $\bm A(\tau)=A(\cos(\Omega \tau), \sin(\Omega \tau),0)$, the phase $\phi$ is given by,
\begin{align}
\label{eq:phase}
\phi= 
\begin{cases}
\frac{\pi}{2} & (\bm r=\hat{\bm x})\\
0 & (\bm r=\hat{\bm y},\hat{\bm z}).
\end{cases}
\end{align}

We obtain the formula of $\varepsilon_{\bm k}^{(n)}$ as,
\begin{widetext}
\begin{enumerate}
\item[1.] Linearly polarized light $\bm E$$\parallel$[100]
with $\bm A(\tau)=A\sin(\Omega \tau)(1,0,0)$\\
\hspace{0.5cm}
\begin{align}
\label{eq:epskn1}
\varepsilon_{\bm k}^{(n)}=\left\{
\begin{array}{ll}
-2t[J_n(A)\cos k_x +\delta_{n0}(\cos k_y+\cos k_z)] & (n\in{\rm even})\\
2itJ_n(A)\sin k_x & (n \in{\rm odd}).
\end{array}
\right.
\end{align}
%
\item[2.] Linearly polarized light $\bm E$$\parallel$[110]
with $\bm A(\tau)=A\sin(\Omega \tau)(1,1,0)$\\
\hspace{0.5cm}
\begin{align}
\label{eq:epskn2}
\varepsilon_{\bm k}^{(n)}=\left\{
\begin{array}{ll}
-2t[J_n(A)(\cos k_x+\cos k_y)+\delta_{n0}\cos k_z] & (n\in{\rm even})\\
2itJ_n(A)(\sin k_x+\sin k_y) & (n\in{\rm odd}).
\end{array}
\right.
\end{align}
%
\item[3.] Circularly polarized light
with $\bm A(\tau)=A(\cos(\Omega \tau),\sin(\Omega t),0)$\\
\hspace{0.5cm}
\begin{align}
\label{eq:epskn3}
\varepsilon_{\bm k}^{(n)}=\left\{
\begin{array}{ll}
-2 t \left[J_n(A)(\cos k_x +\cos k_y)+\delta_{n0}\cos k_z\right] & ({\rm mod}(n,4)=0)\\
-2 t J_n(A)(-\sin k_x +i\sin k_y) & ({\rm mod}(n,4)=1)\\
-2 t J_n(A)(-\cos k_x +\cos k_y) & ({\rm mod}(n,4)=2)\\
-2 t J_n(A)( \sin k_x +i\sin k_y) & ({\rm mod}(n,4)=3).
\end{array}
\right.
\end{align}
\end{enumerate}
\end{widetext}
Here $\varepsilon_{\bm k}^{(0)}$ gives the time-averaged effective band energies. Because the absolute value of the Bessel function is always less than unity, i.e., $|J_0(x)|\le 1$, the bandwidth of the photodriven system is reduced as compared with that in the static system. This photoinduced contraction of the bandwidth is called the dynamical localization effect~\cite{Dunlap1986}.

\subsection{Floquet Green's function}
The Floquet theory is also applicable to the Green’s function method. In the time-periodic system with a time periodicity of $T_{\rm p}$, the Green's function $G$ satisfies the following relation,
\begin{align}
\label{eq:Gfnc1}
G(\tau+T_{\rm p},\tau'+T_{\rm p})=G(\tau,\tau').
\end{align}
We apply the Floquet theory to this two-time Green's function as,
\begin{align}
\label{eq:Gfnc2}
&G_{mn}(\omega)=\int_{-T_{\rm p}/2}^{T_{\rm p}/2} \frac{d\tau_{\rm a}}{T_{\rm p}}
\int_{-\infty}^\infty\;d\tau_{\rm r} e^{i(\omega+m\Omega)\tau-i(\omega+n\Omega)\tau'} G(\tau,\tau'),
\end{align}
where
\begin{align}
\label{eq:Gfnc2a}
\tau_{\rm a}=\frac{\tau+\tau'}{2},
\quad\quad
\tau_{\rm r}=\tau-\tau'.
\end{align}
The following relation holds for the retarded and advanced Green's functions, $G^{\rm R}$, $G^{\rm A}$,
\begin{align}
\label{eq:Gfnc4}
G^{\rm A}_{mn}(\omega)=[G^{\rm R}_{mn}(\omega)]^*.
\end{align}

\subsection{Self-energy of magnons}
Next we calculate the magnon self-energy in the photodriven Kondo-lattice mode by applying the above introduced Floquet Green's function formalism~\cite{Ono2018}. 
To apply this theoretical formalism, the system is required to be steady and time-periodic, even under continuous energy supply from light irradiation. Such a situation can be realized through coupling to a heat bath, with the full Hamiltonian given by,
\begin{equation}
\label{eq:H}
\mathcal{H}=\mathcal{H}_{\rm sys}+\mathcal{H}_{\rm int}+\mathcal{H}_{\rm bath}.
\end{equation}
Here the first term $\mathcal{H}_{\rm sys}$ is the Hamiltonian of the Kondo-lattice model,  the second term $\mathcal{H}_{\rm int}$ describes the coupling between the system and the heat bath, and the last term $\mathcal{H}_{\rm bath}$ is the Hamiltonian of the heat bath.

We apply the Holstein-Primakoff transformation to the localized spin operators in the term $\mathcal{H}_{\rm sys}$ as,
\begin{align}
\label{eq:HPtr1}
&S_{iz}=S-a_i^\dag a_i,\\
\label{eq:HPtr2}
&S_i^{+}=\sqrt{2S}\sqrt{1-\frac{a_i^\dag a_i}{2S}} \; a_i \approx \sqrt{2S} \; a_i,\\
\label{eq:HPtr3}
&S_i^{-} =\sqrt{2S}a_i^\dag \; \sqrt{1-\frac{a_i^\dag a_i}{2S}} \approx \sqrt{2S} \; a_i^\dag.
\end{align}
With these transformations, the term $\mathcal{H}_{\rm sys}$ is rewritten in the form composed of the noninteracting part $\mathcal{H}_0$ and the electron-magnon interaction parts $\mathcal{V}_1$ and $\mathcal{V}_2$ as,
\begin{align}
\label{eq:HHP0}
\mathcal{H}=\mathcal{H}_0+\mathcal{V}_1+\mathcal{V}_2,
\end{align}
where
\begin{align}
\label{eq:HHP1}
&\mathcal{H}_0=\sum_{\bm k,s} \varepsilon_{\bm k+\bm A(\tau)s} c_{\bm k s}^{\dagger} c_{\bm k s},
\\
\label{eq:HHP2}
&\mathcal{V}_1=
\frac{J}{2S N}\sum_{\bm k, \bm k', \bm q, \bm q', s}\delta_{\bm k+\bm q, \bm k'+\bm q'}
\; \text{sgn}(s) \; c_{\bm k s}^{\dagger} c_{\bm k' s}a_{\bm q}^{\dagger} a_{\bm q'},
\\
\label{eq:HHP3}
&\mathcal{V}_2=
-\frac{J}{2}\sqrt{\frac{1}{S N}} \sum_{\bm k, \bm q} 
\left(c_{\bm k \uparrow}^{\dagger} c_{\bm k+\bm q \downarrow} 
a_{\bm q}^{\dagger}+c_{\bm k + \bm q \downarrow}^{\dagger} c_{\bm k\uparrow} a_{\bm q}\right)\notag
\\
&\hspace{1pc} =-\frac{J}{2}\sqrt{\frac{1}{SN}} \sum_{\bm q} 
\left(s_{\bm q}^{+} a_{\bm q}^{\dagger}+ s_{-\bm q}^{-}a_{\bm q}\right).
\end{align}
with
\begin{align}
\label{eq:HHP4}
&\varepsilon_{\bm k s} \equiv \varepsilon_{\bm k}-\frac{J}{2}\text{sgn}(s),
\\
\label{eq:HHP5}
&s_{\bm q}^{+} \equiv \sum_{\bm k} c_{\bm k \uparrow}^{\dagger} c_{\bm k+\bm q \downarrow},
\\
\label{eq:HHP6}
&s_{-\bm q}^{-} \equiv \sum_{\bm k} c_{\bm k + \bm q \downarrow}^{\dagger} c_{\bm k\uparrow}.
\end{align}

We define the following four path-ordered Green's functions on the Konstantinov-Perel' contour,
\begin{align}
  \label{eq: Green function}
  &G_{\bm{k} s, \bm{k}' s'}(\tau, \tau')=-i\text{Tr}\left[\hat{\rho}\;
  \mathcal{T}_C c_{\bm{k} s}(\tau) \; c_{\bm{k}' s'}^{\dagger}(\tau')
  \right],
  \\
  \label{eq: Magnon Green function}
  &D_{\bm{q}, \bm{q}'}(\tau, \tau')=-i\text{Tr}\left[\hat{\rho}\;
  \mathcal{T}_C  a_{\bm{q}}(\tau)\; a_{\bm{q}'}^{\dagger}(\tau')
  \right],
  \\
  \label{eq: Free Green Function}
  &g_{\bm{k} s, \bm{k}' s'}(\tau, \tau')=-i\text{Tr}\left[\hat{\rho}\;
  \mathcal{T}_C c_{I\bm{k} s}(\tau) \; c_{I\bm{k}' s'}^{\dagger}(\tau')
  \right],
  \\
  \label{eq: Free Magnon Green function}
  &d_{\bm{q}, \bm{q}'}(\tau, \tau')=
  -i\text{Tr}\left[\hat{\rho}\;
  \mathcal{T}_C  a_{\bm{q}}(\tau) \; a_{\bm{q}'}^{\dagger}(\tau')
  \right],
\end{align}
where $G_{\bm k s, \bm k' s'}$ ($D_{\bm q, \bm q'}$) is the Green's function of electrons (magnons), and $g_{\bm k s, \bm k' s'}$ ($d_{\bm q, \bm q'}$) is the noninteracting Green's function of electrons (magnons). Here $\hat{\rho}$ is the initial density matrix of the system, and $\mathcal{T}_C$ is the path-ordering operator on the contour. The symbols $c_{\bm k s}(\tau)$ and $c_{\text{I} \bm k s}(\tau)$ are the Heisenberg representation and the interaction representation of the electron operator $c_{\bm k s}$, respectively. 

In the following, we assume that light irradiation begins at time $\tau_{\rm o} \rightarrow -\infty$ and that the system evolves into a time-periodic nonequilibrium steady state sufficiently slowly. Under this assumption, the Konstantinov-Perel' contour is reduced to time-ordered and anti-time-ordered paths along the real-time axis, which allows the treatment based on the Floquet-Keldysh formalism. It should be noted that the real times $\tau$ and $\tau'$ that appear in the following formulation are those measured from a certain moment after the system has settled into the time-periodic steady state and thus differ from the initial time $\tau_0$ used in the numerical simulations. Although this setup differs from that used in the numerical simulations where light irradiation is turned on instantaneously, the present argument is expected to offer valuable insights into the real-time dynamics of the light-irradiated Kondo-lattice model.

For the noninteracting Green's function of electrons $g_{\bm k s, \bm k' s'}(\tau,\tau')$, we consider the one-particle wavefunction $\ket{\Psi(\tau)}$, which satisfies the time-dependent Schr\"odinger equation for the noninteracting Hamiltonian $\mathcal{H}_0$, 
\begin{align}
\label{eq:tScheq2}
\left(i\frac{d}{d\tau}-\mathcal{H}_0\right)\ket{\Psi(\tau)}=0.
\end{align}
Since the Hamiltonian $\mathcal{H}_0$ is given in the diagonalized form with the momentum basis ${\phi_{\bm k s}}\equiv\{c_{\bm k s}^{\dagger}\ket{0}\}$ where $\bm k \in \text{BZ}$ and $s=\uparrow,\downarrow$, the eigenket of this equation should be $\ket{\Psi_{\bm k s}(\tau)}$, and the wavefunction is given by,
\begin{align}
\label{eq:Psikst}
\Psi_{\bm k s}(\tau)
&\equiv\braket{\phi_{\bm k' s'}|\Psi_{\bm{k}s}(\tau)} \notag\\
&=\delta_{\bm k \bm k'}\delta_{ss'}\exp\left[-i\int_{0}^{\tau}d\bar{\tau} \varepsilon_{\bm k s}(\bar{\tau})\right].
\end{align}
We then introduce time-periodic kets $\ket{u_{\bm k s}(\tau)}$ that satisfy,
\begin{align}
\label{eq:ukt}
\ket{\Psi_{\bm k s}(\tau)}
&=\exp \left[-i\varepsilon_{\bm k s}^{(0)}\tau\right]\ket{u_{\bm k s}(\tau)} \notag \\
&=\exp \left[-i\varepsilon_{\bm k s}^{(0)}\tau\right]\sum_{n=-\infty}^{\infty}\ket{u_{\bm{k}s}^{(n)}}e^{-in\Omega t},
\end{align}
where we define,
\begin{align}
\label{eq:eiks0}
\varepsilon_{\bm k s}^{(0)} 
&\equiv \varepsilon_{\bm k}^{(0)}-\frac{J}{2}\text{sgn}(s).
\end{align}
We also introduce the Fourier coefficients $u_{\bm k s}^{(n)}$ as,
\begin{align}
\label{eq:ukn}
u_{\bm k s}^{(n)}
&=\braket{\phi_{\bm k s}|u_{\bm{k}s}^{(n)}}\notag\\
&=\int_{-T_{\rm p}/2}^{T_{\rm p}/2}\frac{d\tau}{T_{\rm p}} e^{in\Omega \tau}\braket{\phi_{\bm k s}|u_{\bm ks}(\tau)}.
\end{align}
As can be seen from the first equation in Eq.~\eqref{eq:ukn}, the newly introduced kets $\ket{u_{\bm k s}^{(n)}}$, which represent the Floquet-Bloch states, are not normalized. This lack of normalization ensures the convergence of the expanded form of $\ket{u_{\bm ks}(\tau)}$.
The retarded noninteracting Green's function $g_{\bm k s, \bm k' s'}^{\rm R}(\tau,\tau')$ of electrons is given by,
\begin{align}
\label{eq:gRt}
g_{\bm k s,\bm k's'}^{\rm R} (\tau,\tau')
&=-i\delta_{\bm k \bm k'}\delta_{ss'}\theta(\tau-\tau')\Psi_{\bm k s}(\tau)\Psi_{\bm k' s'}^*(\tau') \notag\\
&\equiv \delta_{\bm k \bm k'}\delta_{ss'}g_{\bm k s}^{\rm R}(\tau,\tau').
\end{align}
On the other hand, the Floquet representation of this equation is given by,
\begin{align}
\label{eq:gkFlqt}
\bm g_{\bm k s}^{\rm R}(\omega)
=\bm u_{\bm k s} \bm Q_{\bm k s}(\omega) \bm u_{\bm k s}^\dag.
\end{align}
The matrix components of $\bm u_{\bm k s}$ and $\bm Q_{\bm k s}(\omega)$ are, respectively, given by,
\begin{align}
\label{eq:uksmn}
&u_{\bm k s, mn} = u_{\bm k s}^{(m-n)}, \\
\label{eq:Qkmn}
&Q_{\bm k s,mn}(\omega)=
\frac{\delta_{mn}}{\omega+n\Omega-\xi_{\bm k s}^{(0)}+i\eta},
\end{align}
where $\eta$ is a positive infinitesimal. 
\begin{widetext}
Using Eqs.~\eqref{eq:gkFlqt}, the inverse of the noninteracting Green's function is derived as,
\begin{align}
\label{eq:gRinv}
[g_{\bm k s}^{\rm R}(\omega)]^{-1}_{mn}
=\delta_{mn}\left(\omega+n\Omega+\frac{J}{2}\text{sgn}(s)
+\mu+i\eta\right)-\varepsilon_{\bm k}^{(m-n)}.
\end{align}

We now consider Eq.\eqref{eq:HHP0} again to incorporate the coupling between the system and the heat bath into the one-particle Green's function of electrons. We introduce the dissipative one-particle Green's function $\hat{\mathcal{G}}$ as,
\begin{align}
\label{eq:opG}
&\hat{\mathcal{G}}^{-1}(\omega)
=\hat{\bm g}^{-1}(\omega)-\hat{\bm \Sigma}_{\rm bath}(\omega)\notag\\
\Leftrightarrow &
\begin{pmatrix}
\bm{\mathcal{G}}^{\rm R}(\omega) & \bm{\mathcal{G}}^{\rm K}(\omega)\\
0 & \bm{\mathcal{G}}^{\rm A}(\omega)
\end{pmatrix}^{-1}
=
\begin{pmatrix}
[\bm g^{\rm R}(\omega)]^{-1} & 0\\
0 & [\bm g^{\rm A}(\omega)]^{-1}
\end{pmatrix}
-
\begin{pmatrix}
\bm \Sigma_{\rm bath}^{\rm R}(\omega) & \bm \Sigma_{\rm bath}^{\rm K}(\omega)\\
0 & \bm \Sigma_{\rm bath}^{\rm A}(\omega) 
\end{pmatrix},
\end{align}
where
\begin{align}
\label{eq:Dyson2}
\hat{\bm \Sigma}_{\rm bath}(\omega) \equiv 
\begin{pmatrix}
\bm \Sigma_{\rm bath}^{\rm R}(\omega) & \bm \Sigma_{\rm bath}^{\rm K}(\omega)\\
0 & \bm \Sigma_{\rm bath}^{\rm A}(\omega) 
\end{pmatrix}
=
\begin{pmatrix}
-i\Gamma_c\bm I & -2i\Gamma_c \bm F(\omega)\\
0 & i\Gamma_c\bm I
\end{pmatrix},
\end{align}
and
\begin{align}
\label{eq:Dyson3}
F_{mn}(\omega)=\{1-2f(\omega+n\Omega-\mu)\}\delta_{mn}.
\end{align}
Here $\hat{\bm G}$ and $\hat{\bm g}$ denote infinite-dimensional matrices of the Floquet Green's functions for the entire system and the free electron, respectively. $\Gamma_c$ is a spectral function of the heat bath that is coupled with the electrons, $f(\omega)$ is the Fermi-Dirac distribution function, and $\mu$ is the chemical potential of the electrons. We assume a particle reservoir for the heat bath, which has a flat density of states and thus enables us to omit its $\omega$ dependence.
By using Eq.~\eqref{eq:opG} and the relation \eqref{eq:Gfnc4}, we obtain $\bm{\mathcal{G}}^{\rm R}, \bm{\mathcal{G}}^{\rm A}, \bm{\mathcal{G}}^{\rm K}$, and, $\bm{\mathcal{G}}^<$, which are the retarded, advanced, Keldysh, and lesser components of the dissipative one-particle  Green's function, respectively.  
We can also apply the same reasoning to both the one-particle and the dissipative one-particle Green's functions of magnons, $\hat{\bm{d}}$ and $\hat{\bm{\mathcal{D}}}$. Since magnon-related terms are absent in $\mathcal{H}_0$, the inverse of the retarded component is given by,
\begin{align}
\label{eq:Dqw}
[\mathcal{D}_{\bm q}^{\rm R}(\omega)]_{mn}^{-1}=\omega + n\Omega +i\Gamma_a,
\end{align}
where $\Gamma_a$ denotes the spectral function of another bath coupled to the magnon quasiparticles.

\begin{figure}[tb]
\centering
\includegraphics[width = 0.35\columnwidth]{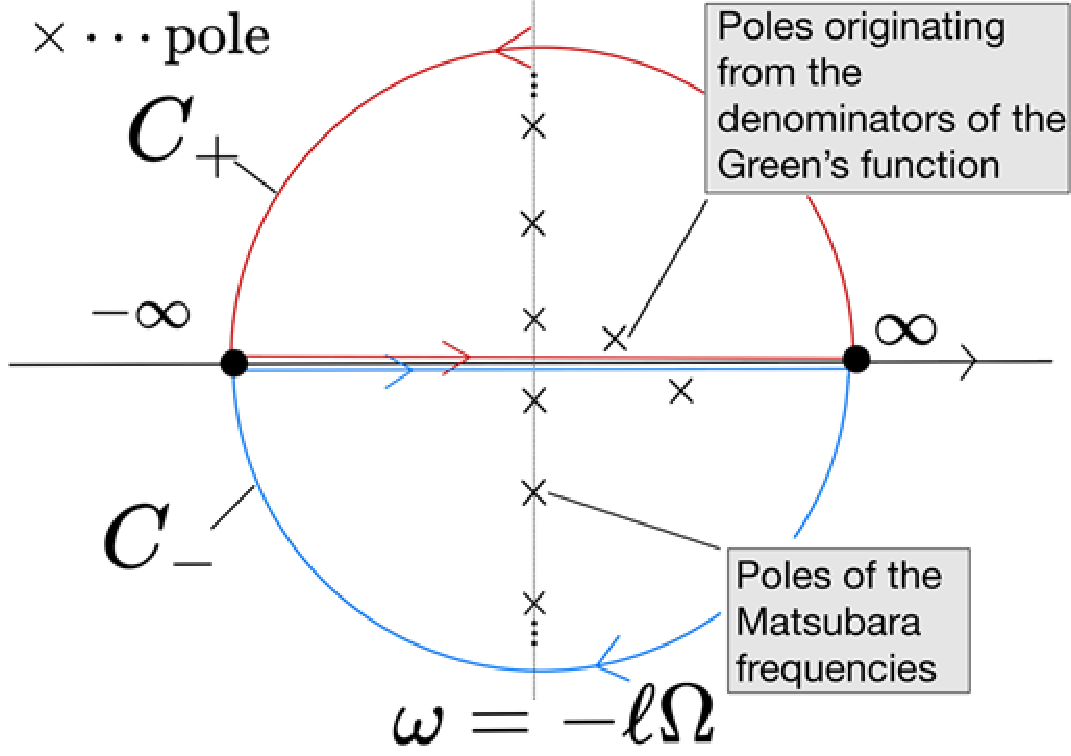}
\caption{Integration paths $C_{\pm}$.}
\label{Fig13}
\end{figure}
In Ref.~\cite{Ono2018}, the self-energies of magnons are calculated based on an assumption that the electrons are nearly noninteracting particles and only contribute to the dissipation, by which the magnon-magnon interactions are mediated. According to this assumption, we expand the S matrix up to the first order of $\mathcal{V}_1$ and the second order of $\mathcal{V}_2$ to obtain the self-energies in the Floquet representation as,
\begin{align}
\label{eq:SgmRa}
\Sigma_{\bm q,mn}^{\rm R}(\omega)
&=\Sigma_{mn}^{\rm R,1}+\Sigma_{\bm q,mn}^{\rm R,2}(\omega)
\notag\\
&=-\frac{iJ}{2SN}\sum_{\bm k,s} \text{sgn}(s) \int_{-\infty}^\infty \frac{d\bar{\omega}}{2\pi}
\mathcal{G}_{\bm k s,(m-n)0}^{<}(\bar{\omega})
\notag\\
&
-\frac{iJ^2}{4SN}\sum_{\bm k, j}\int_{-\infty}^\infty \frac{d\bar{\omega}}{2\pi}
\left\{
\mathcal{G}_{\bm k+\bm q \downarrow,m(n+j)}^{\rm R}(\omega+\bar{\omega}) \;
\mathcal{G}_{\bm k \uparrow,j0}^{\rm K}(\bar{\omega})
+
\mathcal{G}_{\bm k+\bm q \downarrow,m(n+j)}^{\rm K}(\omega+\bar{\omega}) \;
\mathcal{G}_{\bm k \uparrow,j0}^{\rm A}(\bar{\omega})
\right\}.
\end{align}
By performing the residue calculations with integration paths $C_{+}$ and $C_{-}$ shown in Fig.~\ref{Fig13}, the retarded magnon self-energy reads,
\begin{align}
\label{eq:SgmRb}
\Sigma_{\bm q,mn}^{\rm R}(\omega)=\Sigma_{mn}^{\rm R,1}+
\Sigma_{\bm q,mn}^{\rm R,2,L}(\omega)+\Sigma_{\bm q,mn}^{\rm R,2,c1}(\omega) +\Sigma_{\bm q,mn}^{\rm R,2,c2}(\omega),
\end{align}
with
\begin{align}
\label{eq:SgmR1}
&\Sigma_{mn}^{\rm R,1}
=\frac{J}{2SN}\frac{2i\Gamma_c}{(m-n)\Omega+2i\Gamma_c}\sum_{\bm k}
\left[n_{\bm k \uparrow}^{(m-n)}-n_{\bm k \downarrow}^{(m-n)}\right],
\\
\label{eq:SgmR2L}
&\Sigma_{\bm q,mn}^{\rm R,2,L}(\omega)
=\frac{J^2}{2SN}\sum_{\bm k}\sum_{\ell_1,\ell_2} 
\frac{u_{\bm k +\bm q\downarrow, \bm k\uparrow}^{(m+\ell_1)}\; u_{\bm k +\bm q\downarrow, \bm k\uparrow}^{(n+\ell_1-\ell_2)*}\; n_{\bm k\uparrow}^{(\ell_2)}
- u_{\bm k +\bm q\downarrow, \bm k\uparrow}^{(m+\ell_1-\ell_2)}\;
u_{\bm k +\bm q\downarrow, \bm k\uparrow}^{(n+\ell_1)*}\;
n_{\bm k +\bm q\downarrow}^{(-\ell_2)*}}{\omega - \ell_1\Omega - (\varepsilon_{\bm k +\bm q\downarrow}^{(0)}-\varepsilon_{\bm k\uparrow}^{(0)})}
\frac{2i\Gamma_c}{\ell_2\Omega+2i\Gamma_c},
\\
\label{eq:SgmR2c1}
&\Sigma_{\bm q,mn}^{\rm R,2,c1}(\omega)
=-\frac{J^2}{2SN}\sum_{\bm k}\sum_{\ell_1,\ell_2,\ell_3}
\frac{u_{\bm k +\bm q\downarrow, \bm k\uparrow}^{(m+\ell_1)}\; 
u_{\bm k +\bm q\downarrow, \bm k\uparrow}^{(n+\ell_1-\ell_2)*}\; 
u_{\bm k \uparrow}^{(\ell_2 + \ell_3)}\;
u_{\bm k \uparrow}^{(\ell_3)}\;
}{\omega - \ell_1\Omega - (\varepsilon_{\bm k +\bm q\downarrow}^{(0)}-\varepsilon_{\bm k\uparrow}^{(0)})}\notag
\\
&\hspace{10pc}\times
\frac{i\Gamma_c
+\frac{\Gamma_c}{\pi}\left\{\psi\left(\frac{1}{2}-i\beta\; \frac{\xi_{\bm k \uparrow}^{(0)} +(\ell_2 + \ell_3)\Omega+i\Gamma_c}{2\pi}\right)
-\psi\left(\frac{1}{2}+i\beta\; \frac{-\omega+\xi_{\bm k + \bm q \downarrow}^{(0)}+(\ell_1+\ell_3)\Omega-i\Gamma_c}{2\pi}\right)\right\}}
{\omega - (\ell_1-\ell_2)\Omega - (\varepsilon_{\bm k +\bm q\downarrow}^{(0)}-\varepsilon_{\bm k\uparrow}^{(0)})+2i\Gamma_c},
\\
\label{eq:SgmR2c2}
&\Sigma_{\bm q,mn}^{\rm R,2,c2}(\omega)
=\frac{J^2}{2SN}\sum_{\bm k}\sum_{\ell_1,\ell_2,\ell_3}
\frac{u_{\bm k +\bm q\downarrow, \bm k\uparrow}^{(m+\ell_1-\ell_2)}\; 
u_{\bm k +\bm q\downarrow, \bm k\uparrow}^{(n+\ell_1)*}\; 
u_{\bm k + \bm q \downarrow}^{(-\ell_2 + \ell_3)}\;
u_{\bm k + \bm q \downarrow}^{(\ell_3)}\;
}{\omega - \ell_1\Omega - (\varepsilon_{\bm k +\bm q\downarrow}^{(0)}-\varepsilon_{\bm k\uparrow}^{(0)})}\notag
\\
&\hspace{9.5pc}\times
\frac{i\Gamma_c
-\frac{\Gamma_c}{\pi}\left\{\psi^*\left(\frac{1}{2}-i\beta\; \frac{\xi_{\bm k \uparrow}^{(0)} +(\ell_3-\ell_2)\Omega+i\Gamma_c}{2\pi}\right)
-\psi^*\left(\frac{1}{2}+i\beta\; \frac{\omega+\xi_{\bm k + \bm q \downarrow}^{(0)}+(\ell_3-\ell_1)\Omega-i\Gamma_c}{2\pi}\right)\right\}}
{\omega - (\ell_1-\ell_2)\Omega - (\varepsilon_{\bm k +\bm q\downarrow}^{(0)}-\varepsilon_{\bm k\uparrow}^{(0)})+2i\Gamma_c},
\end{align}
where $\psi(z)$ is the digamma function, $\psi^*(z)\equiv\psi(z)^*$ and, 
\begin{align}
 &u_{\bm k + \bm q \downarrow, \bm k \uparrow}^{(m)}\equiv\sum_{\ell}u_{\bm k + \bm q \downarrow}^{(m+\ell)} u_{\bm k \uparrow}^{(\ell)*}
  =\sum_{\ell}\bra{u_{\bm k + \bm q \downarrow}^{(m+\ell)}}
  s_{-\bm q}^{-}
  \ket{u_{\bm k \uparrow}^{(\ell)}},
  \\
  \label{eq:nks}
  &n_{\bm k s}^{(m)} \equiv
\sum_{\ell}u_{\bm k s}^{(m+\ell)} u_{\bm k s}^{(\ell)*}
\left[\frac{1}{2}+\frac{1}{2\pi i}\left\{\psi\left(\frac{1}{2}-i\beta\; \frac{\xi_{\bm k s}^{(0)} +(m + \ell)\Omega+i\Gamma_c}{2\pi}\right)
-\psi\left(\frac{1}{2}+i\beta\; \frac{\xi_{\bm k s}^{(0)} +\ell\Omega-i\Gamma_c}{2\pi}\right)\right\}\right].
\end{align}
Here the wave vector $\bm k$ runs over the first Brillouin zone, $\ell_1,\ell_2$, and $\ell_3$ takes all integers from $-\infty$ to $\infty$, and $\beta(\equiv 1/T)$ is the inverse temperature. $\xi_{\bm k s}^{(0)}$ is defined as, 
\begin{align}
\label{eq:xiks0}
\xi_{\bm k s}^{(0)}\equiv \varepsilon_{\bm k s}^{(0)}-\mu,
\end{align}
where $\mu$ is the chemical potential. In the calculations, we assume $(2n+1)\pi/\beta \ne \Gamma_c$. An identity of the digamma function that holds for $\omega \in \mathbb{C}$ is given as follows,
\begin{align}
\label{eq:rflctfrml}
\frac{1}{2}+\frac{1}{2\pi i}\left[\psi\left(\frac{1}{2}-\frac{i\beta\omega}{2\pi}\right)-\psi\left(\frac{1}{2}+\frac{i\beta\omega}{2\pi}\right)\right]= f(\omega).
\end{align}
Each of $\Sigma_{\bm q, mn}^{\rm R,2,L}$, $\Sigma_{\bm q,mn}^{\rm R,2,c1}$, and $\Sigma_{\bm q, mn}^{\rm R,2,c2}$ contains terms that diverge at  $\omega = \ell_1\Omega + (\varepsilon_{\bm k +\bm q\downarrow}^{(0)}-\varepsilon_{\bm k\uparrow}^{(0)})$. However, when summed, the divergences cancel out, and the total expression converges to the derivatives of the second factors of $\Sigma_{\bm q,mn}^{\rm R,2,c1}$, and $\Sigma_{\bm q, mn}^{\rm R,2,c2}$ at $\omega = \ell_1\Omega + (\varepsilon_{\bm k +\bm q\downarrow}^{(0)}-\varepsilon_{\bm k\uparrow}^{(0)})$. To clarify the photoinduced magnetic phase transitions observed in our numerical simulations, we examine the parameters used in the simulations, that is, $J=14t$ for the Kondo coupling constant and $\Omega=0.5t$ for the light frequency. We also take $\beta = 10000/t$, and $\Gamma_c$=0.02.

Now let us discuss the properties of each term to validate the derived formulas. The first term $\Sigma_{mn}^{\rm R, 1}$ corresponds to the Zeeman energy associated with effective magnetic fields generated by electrons which act on the localized spins. We can see this aspect by considering the expression of $\Sigma_{mn}^{\rm R,1d}$ in Eq.~\eqref{eq:SgmR1} at the equilibrium state with $\bm A=0$, and $\Gamma_c\rightarrow +0$, for which the relations $u_{\bm k s}^{(m)}=\delta_{m0}$, $\xi_{\bm k s}^{(0)}=\xi_{\bm ks}$, and $n_{\bm k s}^{(\ell)}=f(\xi_{\bm k s})\delta_{\ell 0}$ hold. Substituting these relations to Eq.~\eqref{eq:SgmR1}, we indeed obtain the expression of $\Sigma_{mn}^{\rm R,1d}$ at equilibrium as,
\begin{align}
\label{eq:SgmR1db}
\Sigma^{\rm R,1d}_{mn}
=\frac{J\delta_{mn}}{2SN}\sum_{\bm k}(f(\xi_{\bm k \uparrow})-f(\xi_{\bm k \downarrow})).
\end{align}
In the nonequilibrium regime, the effect of electron excitations induced by a periodic drive is incorporated into $\Sigma_{mn}^{\rm R,1}$ by replacing the Fermi-Dirac distribution with the nonequilibrium one-particle distribution function $n_{\bm k s}^{(m)}$. This distribution function becomes time-independent in the limit of weak system-bath coupling, and it is denoted as $n_{\bm ks}^{(0)}$. This is because the nonequilibrium distribution function $n_{\bm ks}$ does not account for the effects of magnon perturbations, and the only source of the time dependence is the coupling to a heat bath. We note that the same expression was derived for the distribution function of excited electrons in Ref.~\cite{Matsyshyn2023}, using a periodically driven noninteracting system coupled to a heat bath with a flat density of states. However, no previous work has analytically derived the expression within the Floquet Green's function formalism.

We next discuss the term $\bm \Sigma_{\bm q}^{\rm R,2}(\tau,\tau')$, which is proportional to the nonequilibrium dynamical magnetic susceptibility of the electron system $\chi_{\bm q}^{\rm R}(\tau,\tau')$. The susceptibility $\chi_{\bm q}(\tau,\tau')$ is defined on the Konstantinov-Perel' contour as,
\begin{align}
\label{eq:chiq}
\chi_{\bm q}(\tau,\tau')=\frac{2i}{N}\rm Tr \it\left[\hat{\rho}\; \mathcal{T}_C s_{\bm{q}}^{+}(\tau) (s_{\bm{q}}^{+})^\dag(\tau') \right].
\end{align}
The relation between $\chi_{\bm q}(\tau,\tau')$ and $\Sigma_{\bm q}^{\rm R,2}(\tau,\tau')$ is given by,
\begin{align}
\label{eq:chiSgm}
\Sigma_{\bm q}^{\rm R,2}(\tau,\tau')=-\frac{J^2}{4S}\chi_{\bm q}(\tau,\tau').
\end{align}
We compare the formula of $\chi_{\bm q, mn}^{\rm R}(\omega)$ for the equilibrium system with that for the nonequilibrium system coupled  weakly to the heat bath to investigate possible photoinduced effects . At equilibrium, the dynamical magnetic susceptibility is given by,
\begin{align}
\label{eq:chiReq}
\chi_{\bm q, mn}^{\rm R, eq}(\omega) 
=-\frac{2\delta_{mn}}{N}\sum_{\bm k}\frac{f(\xi_{\bm k \uparrow})-f(\xi_{\bm k + \bm q \downarrow})}{\omega +m\Omega -(\varepsilon_{\bm k+\bm q \downarrow}-\varepsilon_{\bm k \uparrow})+i\eta}.
\end{align}
On othe other hand, the susceptibility for the nonequilibrium system is given by,
\begin{align}
\label{eq:chiRwbc}
  \chi_{\bm q, mn}^{\rm R, neq}(\omega)
  =-\frac{2}{N}\sum_{\bm k}\sum_{\ell_1}\frac{
  \left|\bra{u_{\bm k + \bm q \downarrow}^{(m+\ell_1)}}
  s_{-\bm q}^{-}
  \ket{u_{\bm k\uparrow}^{(n+\ell_1)}}\right|^2}
  {\omega-\ell_1\Omega-(\varepsilon_{\bm k+\bm q \downarrow}^{(0)}-\varepsilon_{\bm k \uparrow}^{(0)})+2i\Gamma_c}\left[n_{\bm k \uparrow}^{(0)}- n_{\bm k + \bm q\downarrow}^{(0)}\right].
\end{align}
Here we consider the weak bath-coupling limit with $\Gamma_c/\Omega \ll 1$. Contributions from processes that combine individual electron excitations and photoinduced $(m-n)$-photon absorptions are incorporated to $\chi_{\bm q, mn}^{\rm R, neq}(\omega)$ via the transition probability between the two Floquet-Bloch states $\ket{u_{\bm k \uparrow}^{(m+\ell_1)}}$ and $\ket{u_{\bm k+ \bm q \downarrow}^{(n+\ell_1)}}$. Effects of changes in the electron occupation are also taken into account through the nonequilibrium distribution function.

It is important to note that the susceptibility in Eq.~\eqref{eq:chiRwbc} takes the same form as the Floquet-Kubo formula derived in Ref.~\cite{Rudner2020}. While the original Floquet-Kubo formula is independent of any specific one-particle distribution function and applies to arbitrary time-independent distributions, the susceptibility which we have derived depends on the nonequilibrium distribution function $n_{\bm k s}^{(0)}$ in Eq.~\eqref{eq:nks}. We, therefore, conclude that $\chi_{\bm q,mn}^{\rm R, neq}(\omega)$ constitutes a variant of the Floquet-Kubo formula, and that the susceptibility, which is proportional to the sum of $\Sigma_{\bm q,mn}^{\rm R,2,L}$, $\Sigma_{\bm q,mn}^{\rm R,2,c1}$ and $\Sigma_{\bm q,mn}^{\rm R,2,c2}$ for a general heat-bath spectral function, represents its extension.

\subsection{Magnon dispersion relation of the time-periodic Kondo-lattice model}
In the previous section, we derived the self-energies of magnons in the Floquet representation and considered its properties. Here we calculate the magnon dispersion relation numerically by solving the following equation, which is equivalent to the eigenvalue equation of the Floquet Hamiltonian,
\begin{align}
\label{eq:deteq}
&\det\left(\text{Re}\left[\left(\bm{\mathcal{D}}_{\bm q}^{\rm R}(\tilde{\omega}_{\bm q}) \right)^{-1}
-\bm \Sigma_{\bm q}^{\rm R}(\tilde{\omega}_{\bm q}) \right] \right)=0
\notag\\
\Leftrightarrow
&\begin{vmatrix}
\ddots && \vdots && \vdots && \vdots && \\ \\
\cdots && \tilde{\omega}_{\bm q}-\Omega-\Sigma_{\bm q,-1-1}^{\rm R}(\tilde{\omega}_{\bm q}) &&
-\Sigma_{\bm q,-10}^{\rm R}(\tilde{\omega}_{\bm q}) && 
-\Sigma_{\bm q,-11}^{\rm R}(\tilde{\omega}_{\bm q}) && \cdots \\ \\
\cdots && -\Sigma_{\bm q,0-1}^{\rm R}(\tilde{\omega}_{\bm q}) && 
\tilde{\omega}_{\bm q}-\Sigma_{\bm q,00}^{\rm R}(\tilde{\omega}_{\bm q}) &&
-\Sigma_{\bm q,01}^{\rm R}(\tilde{\omega}_{\bm q}) && \cdots \\ \\
\cdots && -\Sigma_{\bm q,1-1}^{\rm R}(\tilde{\omega}_{\bm q}) && 
-\Sigma_{\bm q,10}^{\rm R}(\tilde{\omega}_{\bm q})&&
 \tilde{\omega}_{\bm q}+ \Omega-\Sigma_{\bm q,11}^{\rm R}(\tilde{\omega}_{\bm q}) && \cdots \\ \\
  && \vdots && \vdots && \vdots && \ddots\\
\end{vmatrix}
=0
\end{align}
\end{widetext}
It is not easy to solve this eigen equation because the self-energies $\Sigma_{\bm q,mn}$ are also functions of $\tilde{\omega}_{\bm q}$. To simplify this equation, we use a property of $\Sigma_{\bm q,mn}^{\rm R}(\omega)$, which is expressed by,
\begin{align}
\label{eq:SgmReq}
\Sigma_{\bm q,mn}^{\rm R}(\omega)=\Sigma_{\bm q,(m+\ell)(n+\ell)}^{\rm R}(\omega-\ell\Omega).
\end{align}
This relation indicates that when $\tilde{\omega}_{\bm q}$ is an eigenvalue, $\tilde{\omega}_{\bm q}+\ell\Omega$ is also an eigenvalue, and thus we need to find only one eigenvalue to obtain the entire magnon dispersion relation in the Floquet state. Moreover, the approximated value of $\tilde{\omega}_{\bm q}$ is obtained as a solution of the following equation,
\begin{align}
\label{eq:cond1}
&\text{Re} \left[\left[\mathcal{D}_{\bm q,00}^{\rm R}(\omega_{\bm q})\right]^{-1}
-\Sigma_{\bm q,00}^{\rm R}(\omega_{\bm q}) \right]=0
\notag\\
\; \Leftrightarrow \;
&\omega_{\bm q}-\text{Re}\left[\Sigma_{\bm q,00}^{\rm R}(\omega_{\bm q})\right]=0.
\end{align}
This approximation is justified when the absolute values of the nondiagonal terms $\Sigma_{\bm q,m \ne n}$ are sufficiently small as compared to the energy scale considered. In our case, the typical energy scale is governed by the Floquet Brillouin zone $-\Omega/2 \le \omega\le\Omega/2$. Therefore, the validity of this approximation is determined by the following equation,
\begin{align}
\label{eq:cond2}
|\tilde{\omega}_{\bm q} -\omega_{\bm q}|
&\sim \left| \sum_{\ell \ne 0}\frac{|\Sigma_{\bm q,0\ell}^{\rm R}(\omega_{\bm q})|^2}
{\omega_{\bm q}-\omega_{\bm q}-\ell\Omega}\right|
\notag\\
&\sim \left|\frac{|\Sigma_{\bm q,0-1}^{\rm R}(\omega_{\bm q})|^2}{\Omega}
-\frac{|\Sigma_{\bm q,01}^{\rm R}(\omega_{\bm q})|^2}{\Omega}\right|
\notag\\
&\ll \Omega
\end{align}
In the calculations, we first solve Eq.~\eqref{eq:cond1} numericaly to obtain the value of $\omega_{\bm q}$. Using thus obtained $\omega_{\bm q}$, we calculate $\Sigma_{\bm q,0-1}^{\rm R}(\omega_{\bm q})$ and $\Sigma_{\bm q,0,1}^{\rm R}(\omega_{\bm q})$ to check if Eq.~\eqref{eq:cond2} is satisfied. We find  that the maximum absolute values of the nondiagonal terms are 0.05 at most. This fact indicates that the magnon energy can be well approximated by the solution of Eq.~\eqref{eq:cond1}, and the hybridizations among Floquet states with different photon numbers are negligible.

The integer variables $\ell_1$, $\ell_2$ and  $\ell_{3}$ for summations are ranging from $-\infty$ to $\infty$, but for the actual calculations, truncations are necessary. We determine their ranges such that the unitarity of the matrix $\bm u_{\bm k}$ is sufficiently satisfied as,
\begin{align}
\label{eq:cond3}
&^{\forall} |m|\le N_{\rm p} \hspace{1pc} {\rm s.t.} \notag \\
&\left| \sum_{\ell=-N_{\rm p}}^{N_{\rm p}}
u_{\bm k s}^{(m+\ell)} \; u_{\bm ks}^{(\ell)*} - \delta_{m0}\right| 
\le 1.0\times 10^{-6}.
\end{align}
This condition guarantees sufficient accuracy for the calculated self-energies. Typically, $N_{\rm p}$ turns out to be ranging from 4 to 24 depending on the amplitude and the polarization of the light electromagnetic field $\bm A(\tau)$. Under these conditions, Eq.~\eqref{eq:cond1} is solved numerically at each wavevector point $\bm q$ using Newton's method. We start the iterative calculation of the Newton's method by setting $\omega=0$ first and stop it when the solution is converged within an error less than $10^{-4}$. The calculated results are shown in Fig.~\ref{Fig11}.

In the calculated magnon dispersion relations for linearly polarized lights $\bm E$$\parallel$[100] and $\bm E$$\parallel$[110] in Figs.11(a) and (b), we find that softening and inversion of the dispersion relations towards negative excitation energies happen around a momentum parallel to the light polarization $\bm E$. Specifically, the softening occurs at $\bm q=(\pi,0,0)$ (X point) for $\bm E$$\parallel$[100], while it occurs at $\bm q=(\pi,\pi,0)$ (M point) for $\bm E$$\parallel$[110]. Moreover, the magnon excitation energies become negative at these momentum points when the light intensity exceeds a threshold value ($A \gtrsim 1.2$), which indicates instability of the ferromagnetic ground state towards the corresponding AFM state. The calculated magnon dispersion relations for circularly polarized light in Fig.11(c) show similar behaviors to those for the linearly polarized light, especially those for $\bm E$$\parallel$[110] in Fig.11(b).

Such light-polarization dependence of the magnon-band softening and inversion is attributable to an effect of the periodic drive. This can be understood from a comparison between the dispersion relation at equilibrium $\omega_{\bm q}^{\rm eq}$ and that at nonequilibrium $\omega_{\bm q}^{\rm neq}$. From Eq.~\eqref{eq:cond1}, the former is derived as,
\begin{align} 
\label{eq:cond4} 
\omega_{\bm q}^{\rm eq}
&=\frac{J}{2SN}\sum_{\bm k}f(\varepsilon_{\bm k\uparrow}) \frac{\varepsilon_{\bm{k}+\bm{q}}-\varepsilon_{\bm{k}}}{J+\varepsilon_{\bm{k}+\bm{q}}-\varepsilon_{\bm{k}}}.
\end{align}
Here the contribution from $f(\varepsilon_{\bm k + \bm q \downarrow})$ is neglected by considering the up-polarized ferromagnetic state. On the other hand, the latter is derived as,
\begin{align}
\label{eq:cond5} 
\omega_{\bm{q}}^{\rm neq} = & \frac{J}{2SN}\sum_{\bm{k}}n_{\bm{k}\uparrow}^{(0)} \frac{\varepsilon_{\bm{k}+\bm{q}}^{(0)}-\varepsilon_{\bm{k}}^{(0)}}{J+\varepsilon_{\bm{k}+\bm{q}}^{(0)}-\varepsilon_{\bm{k}}^{(0)}} \notag \\
-&\frac{J}{2SN}\sum_{\bm{k}}n_{\bm{k}\uparrow}^{(0)} \frac{J\sigma_{\bm k + \bm q\downarrow, \bm k\uparrow}^2}{(J+\varepsilon_{\bm{k}+\bm{q}}^{(0)}-\varepsilon_{\bm{k}}^{(0)})^3},
\end{align}
where $\sigma_{\bm k + \bm q\downarrow, \bm{k}\uparrow}^2$ denotes the standard deviation of $\varepsilon_{\bm{k}+\bm{q}+\bm{A}(\tau)\downarrow}-\varepsilon_{\bm{k}+\bm{A}(\tau)\uparrow}$, which characterizes the extent of temporal fluctuations of the excitation energy from the state $\ket{u_{\bm k \uparrow}(\tau)}$ to the state $\ket{u_{\bm k + \bm q \downarrow}(\tau)}$. The expression of $\sigma_{\bm k + \bm q\downarrow, \bm{k}\uparrow}^2$ is given by,
\begin{align} 
\label{eq:cond8} 
&\sigma_{\bm k + \bm q\downarrow, \bm{k}\uparrow}^2 \notag\\
=&\int_{0}^{T_p} \frac{d\tau}{T_p}\left(\varepsilon_{\bm{k}+\bm{q}+\bm{A}(\tau)\downarrow}-\varepsilon_{\bm{k}+\bm{A}(\tau)\uparrow}\right)^2 -\left(\varepsilon_{\bm{k}+\bm{q}\downarrow}^{(0)}-\varepsilon_{\bm{k}\uparrow}^{(0)}\right)^2 \notag\\
=&\sum_{\ell = -\infty}^{\infty}u_{\bm{k}+\bm{q}\downarrow,\bm{k}\uparrow}^{(\ell)}\ell^2\Omega^2u_{\bm{k}+\bm{q}\downarrow,\bm{k}\uparrow}^{(\ell)*}.
\end{align}
Note that to drive Eq.~\eqref{eq:cond5} we consider the weak system-bath coupling regime $\Gamma_c/\Omega\ll 1$, which is justified because in real Kondo-lattice magnets, the coupling between the electrons and the bath degrees of freedom are weak enough as compared to the Kondo coupling and the transfer integrals. We expand the right-hand side of Eq.~\eqref{eq:chiRwbc} to derive Eq.~\eqref{eq:cond5} up to the second order of $\delta \equiv \ell_1\Omega/(\varepsilon_{\bm k+\bm q}^{(0)}-\varepsilon_{\bm k}^{(0)}+J)$ by assuming that the dominant contributions come from the terms with $|\delta|\ll 1$.

\begin{figure*}[tb]
\centering
\includegraphics[scale=1.2]{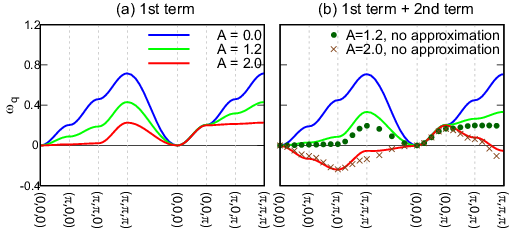}
\caption{(a) Magnon dispersion relations under irradiation with linearly polarized light $\bm E$$\parallel$$[110]$ with $\bm A(\tau) = A(\sin(\Omega\tau),\sin(\Omega\tau),0)$ calculated using only the first term in Eq.~\eqref{eq:cond5}. (b) Magnon dispersion relations calculated using both the first and second terms Eq.~\eqref{eq:cond5} (lines) and those calculated using Eqs.~\eqref{eq:SgmR1}-\eqref{eq:SgmR2c2} for the magnon self-energy (dots and cross symbols).}
\label{Fig14}
\end{figure*}
The expression of $\omega_{\bm{q}}^{\rm neq}$ in Eq.~\eqref{eq:cond5} is composed of two terms. The first term has the same form with $\omega_{\bm q}^{\rm eq}$ in Eq.~\eqref{eq:cond4} upon the substitutions $\varepsilon_{\bm k}\rightarrow\varepsilon_{\bm k}^{(0)}$ and $f(\xi_{\bm k\uparrow})\rightarrow n_{\bm k\uparrow}^{(0)}$. It is revealed that this first term alone cannot account for the softening of magnon dispersions enough significant to realize the negative excitation energy [see also Fig.~\ref{Fig14}]. Indeed, an additional contribution from the second term is required for its realization. Importantly, this second term is unique to the photodriven nonequilibrium system as it contains the temporal fluctuations of time-dependent excitation energy. It induces further softening of the magnon dispersion relations to realize the instability of the ground-state ferromagnetic phase at momentum points corresponding to the low-dimensional AFM correlations. This effect has been overlooked so far and has turned out to play a crucial role for the photoinduced dynamics of long-range magnetic orders.

\section{Appendix: Initial spin configuration}
\label{sec:Appendix3}
For the numerical simulations in the present work, we have started with a FM spin configuration, in which the localized spins are almost polarized in the $z$ direction. However, light irradiation cannot induce spin-orientation changes in a perfectly polarized FM state without numerical rounding errors of numerical simulations. Thus we consider small random deviations from the $z$ direction for the local spin orientations. Specifically, the initial spin configurations are given by,
\begin{align}
\label{eq:Sflc0}
\bm S_i(\tau=0)=(\sin\theta_i \cos\phi_i, \sin\theta_i \sin\phi_i, \cos\theta_i),
\end{align}
with
\begin{align}
\label{eq:Sflc1}
(\theta_i, \phi_i)=(m_i \delta\theta/N, 2\pi n_i/N),
\end{align}
where $m_i$ and $n_i$ are random integer numbers ranging from 1 to $N$. In the real material system, these deviations are caused by thermal fluctuations at finite temperatures. The relationship between the magnitude of $\delta\theta$ and the temperature $T$ can be obtained by the following argument. At equilibrium before the light irradiation, the enegy of the localized spin system per site is given by the exchange coupling with fully polarized electrons, which is given by,
\begin{align}
\label{eq:Sflc2}
&-\frac{1}{N}\sum_i \frac{Jn_{\rm e}}{2}\cos(\theta_i)
\notag\\
&\hspace{0.5cm}
=-\frac{\partial}{\partial \beta} \int_{-1}^1 d(\cos\theta)\exp\left[
-\beta\frac{Jn_{\rm e}}{2}\cos\theta \right]
\notag\\
&\hspace{0.5cm}
=-\frac{Jn_{\rm e}}{2}\coth\left(\frac{ Jn_{\rm e}}{2T}\right) + T
\notag\\
&\hspace{0.5cm}
\approx -\frac{Jn_{\rm e}}{2}\left(1-\frac{2T}{Jn_{\rm e}}\right).
\end{align}
In fact, this energy can be reproduced even if we simply assume that the random polar angles $\theta_i$ are uniformly distributed within a range of $0 \le \theta_i \le \delta\theta$. With this simplification, the enegy of the localized spin system per site is given by,
\begin{align}
\label{eq:Sflc3}
-\frac{1}{N}\sum_i \frac{Jn_{\rm e}}{2}\cos(\theta_i)
=-\frac{Jn_{\rm e}}{2}\left(1-\frac{(\delta\theta)^2}{6}\right).
\end{align}
Therefore, we can mimic the FM spin configuration at finite temperature $T$ by considering the uniformly distributed random spin polar angles $\theta_i$ by setting $\delta\theta$ to satisfy the following relation,
\begin{align}
\label{eq:Sflc4}
\frac{Jn_{\rm e}(\delta\theta)^2}{12}=T.
\end{align}

\section*{Acknowledgements}
The authors thank Atsushi Ono for insightful discussions. This work was supported by JSPS KAKENHI (No.~24H02231 and No.~25H00611), JST CREST (No.~JPMJCR20T1), and a Waseda University Grant for Special Research Projects (No.~2024C-153 and No.~2025C-133).

\end{document}